**Title:** The canonical semantic network supports residual language function in chronic post-stroke aphasia

**Authors:** Joseph C. Griffis[1], Rodolphe Nenert[2], Jane B. Allendorfer[2], Jennifer Vannest[3], Scott Holland[3], Aimee Dietz[4], Jerzy P. Szaflarski[2]

**Institutional Affiliations:** University of Alabama at Birmingham Department of Psychology[1], University of Alabama at Birmingham Department of Neurology[2], Cincinnati Children's Hospital Medical Center, Cincinnati, OH[3], University of Cincinnati Academic Health Center, Cincinnati, OH[4]

**Corresponding Author Information:** Joseph C. Griffis (joegriff@uab.edu)
Department of Neurology and UABEC, University of Alabama at Birmingham, 312 Civitan International Research Center, 1719 6th Avenue South, Birmingham, AL 35294-0021

# Abstract

Current theories of language recovery after stroke are limited by a reliance on small studies. Here, we aimed to test predictions of current theory and resolve inconsistencies regarding right hemispheric contributions to long-term recovery. We first defined the canonical semantic network in 43 healthy controls. Then, in a group of 43 patients with chronic post-stroke aphasia, we tested whether activity in this network predicted performance on measures of semantic comprehension, naming, and fluency while controlling for lesion volume effects. Canonical network activation accounted for 22-33% of the variance in language test scores. Whole-brain analyses corroborated these findings, and revealed a core set of regions showing positive relationships to all language measures. We next evaluated the relationship between activation magnitudes in left and right hemispheric portions of the network, and characterized how right hemispheric activation related to the extent of left hemispheric damage. Activation magnitudes in the each hemispheric network were strongly correlated, but four right frontal regions showed heightened activity in patients with large lesions. Activity in two of these regions (inferior frontal gyrus pars opercularis and supplementary motor area) was associated with better language abilities in patients with larger lesions, but poorer language abilities in patients with smaller lesions. Our results indicate that bilateral language networks support language processing after stroke, and that right hemispheric activations related to extensive left hemisphere damage occur outside of the canonical semantic network and differentially relate to behavior depending on the extent of left hemispheric damage.

# Introduction

## Functional neuroimaging of language recovery after stroke

Aphasia commonly occurs in patients with strokes affecting the left middle cerebral artery (LMCA) territory, and is one of the most debilitating consequences of stroke since language impairments affect nearly every aspect of daily life [Maas et al., 2012]. Language recovery after LMCA stroke is variable, with many survivors experiencing chronic deficits [Charidimou et al., 2014; Lazar et al., 2008; Pedersen, 1995]. Current theories of aphasia recovery are based primarily on evidence obtained from functional neuroimaging studies [Heiss and Thiel, 2006; Saur and Hartwigsen, 2012; Turkeltaub et al., 2011]. During early recovery, patients have been observed to show an up-regulation of right hemispheric responses to language tasks [Heiss et al., 1999; Saur et al., 2006] that is thought to reflect a transient compensatory mechanism triggered by the acute disruption of left hemispheric function by the ischemic event [Hartwigsen et al., 2013; Saur et al., 2006; Thiel et al., 2006]. In later stages of recovery, the gradual reinstatement of left hemispheric capacity for language processing is thought to decrease the need for right hemispheric compensation, resulting in a return of relatively balanced hemispheric activation patterns [Hamilton et al., 2011; Heiss and Thiel, 2006; Saur et al., 2006; Saur and Hartwigsen, 2012; Turkeltaub et al., 2011]. Thus, the general principle to be inferred from functional neuroimaging studies of language recovery after stroke might be summarized as follows: good long-term language outcomes depend primarily on the preservation and/or restoration of close-to-normal function in canonical brain language networks.

While there is consistent support in the literature for a primary role of the canonical left hemispheric language networks in supporting language recovery after stroke, the role of the right hemisphere is less clear, particularly with regard to the chronic recovery phase. This is likely due, in part, to an oversimplified conceptualization of right hemispheric contributions to recovery in much of the literature [Turkeltaub et al., 2012]. Nonetheless, seemingly discrepant findings regarding the role of the right hemisphere pose an obstacle to developing a deeper understanding of how/when the right hemisphere contributes to language recovery. For example, there is evidence that activation of right inferior frontal [Belin et al., 1996; Griffis et al., 2016b; Rosen et al., 2000; Winhuisen et al., 2005; Winhuisen et al., 2007] and right superior temporal [Heiss et al., 1999; Karbe et al., 1998; Szaflarski et al., 2013] areas during language tasks are associated with poorer language function, although there is also evidence that activation of right inferior frontal [Mattioli et al., 2014; van Oers et al., 2010; Raboyeau et al., 2008; Saur et al., 2006] and right anterior superior temporal [Crinion and Price, 2005] regions may support residual language processing and contribute to language recovery.

One explanation for such seemingly discrepant relationships is that when the left hemisphere is not sufficiently preserved, the initial reliance on right hemispheric language network homologues persists beyond early recovery, and the lack of left hemispheric involvement leads to poorer recovery relative to when left hemispheric function is successfully restored [Heiss et al., 1999; Heiss and Thiel, 2006; Karbe et al., 1998; Szaflarski et al., 2013; Turkeltaub et al., 2011]. In this case, the direction of the observed relationship between right hemispheric activation and language task performance might depend on the characteristics of the patient sample. For example, right hemispheric activation might show a positive relationship to language abilities in a sample that primarily consists of severely impaired patients with extensive left hemispheric damage, whereas a negative relationship might be observed in a sample that includes patients with various degrees of impairment and varying extents of left hemispheric damage [Saur and Hartwigsen, 2012].

**Sample sizes in functional neuroimaging studies of aphasia recovery**

A limitation shared by many of the studies that form the foundation for current theories of recovery is that they typically utilize small patient samples [Saur and Hartwigsen, 2012], and this has the potential to (1) increase the risk of detecting spurious effects that do not generalize beyond the sample being studied, (2) reduce power to detect real effects, and (3) produce inflated estimates of the magnitudes of detected effects. Indeed, it has been previously noted in the context of neuroimaging that correlation estimates based on small samples are also unstable and can be disproportionately influenced by outlier data points [Poldrack, 2012]. Further, the power to detect between-subject effects is reduced in small samples [Yarkoni, 2009], and these relationships are often of primary interest in neuroimaging studies of language recovery after stroke. Even for real effects, small sample sizes inflate estimates of effect size, such that for a relationship with a population *r* of 0.3, the average observed sample *r* in samples of 20 patients is expected to be around 0.73 [Yarkoni, 2009].

To illustrate the prevalence of small sample sizes in functional neuroimaging studies (i.e. not including structural/lesion-mapping studies) of post-stroke aphasia, we performed a literature search using Google Scholar. Using the search terms "aphasia fmri", "aphasia pet", "aphasia neuroimaging", "aphasia recovery", "aphasia imaging", and "language recovery after stroke", we identified 84 functional neuroimaging (i.e. not including structural/lesion-mapping studies) studies published between 1995 and 2016 with accessible texts. Across all 84 studies, the average patient sample size was only about 10 patients (mean = 10.33, SD = 8.44; Figure 1), and 87% (73/84) had fewer than 20 patients. The single largest sample was reported by a study published this year [Geranmayeh et al., 2016] that collected data from 53 sub-acute patients (see Supplementary Material S1 for individual study sample sizes and a full reference list). While these estimates are based on a relatively coarse literature search, we note that they are in close agreement with the previous observation by Saur and Hartwigsen [2012] that neuroimaging studies of therapy-induced aphasia recovery typically feature typically less than 10 patients.

Despite the limitations of small samples in individual studies, meta-analytic methods such as activation likelihood estimation (ALE) provide a means for identifying consistent effects in the literature [Wager et al., 2007]. However, we are aware of only one such published meta-analysis of the functional neuroimaging of aphasia recovery literature [Turkeltaub et al., 2011]. Larger studies, in conjunction with meta-analytic assessments of the published literature, are crucial for resolving discrepancies in the literature and for building generalizable theories of recovery that are based on robust empirical evidence.

**Statistical control for lesion volume in functional neuroimaging of aphasia recovery**

A second potential limitation shared by most functional neuroimaging studies of language recovery after stroke is a lack of statistical control for lesion volume effects on imaging-behavior relationships. This is somewhat surprising given that the use of statistical controls to reduce confounds related to lesion volume effects is relatively common in lesion-symptom mapping research [Rorden and Karnath, 2004; Schwartz et al., 2009; Zhang et al., 2014]. The logic behind using statistical controls to account for lesion volume effects is that since (1) larger lesions are often associated with more severe impairments [Allendorfer et al., 2012; Butler et al., 2014; Cheng et al., 2014; Karnath et al., 2004; Kümmerer et al., 2013; Meltzer et al., 2013; van Oers et al., 2010; Rorden and Karnath, 2004; Szaflarski et al., 2013; Yarnell et al., 1976], and (2) larger lesions have a higher probability of including both task-relevant and task-irrelevant voxels [Karnath et al., 2004; Rorden and Karnath, 2004; Schwartz et al., 2009; Zhang et al., 2014], lesion volume effects have the potential to introduce bias into measurements of lesion-behavior relationships at voxels that are primarily damaged in patients with large lesions [Karnath et al., 2004; Zhang et al., 2014]. In functional neuroimaging, the probability that a voxel is active during task performance depends on the probability that the tissue at that voxel is spared, and the probability that the tissue at a given voxel in the left hemisphere is spared depends on the size of the lesion. Therefore, the potential for lesion volume confounds is logically extendable to relationships between functional neuroimaging activation and behavior, and it is possible that a lack of controls for such effects could introduce bias into functional neuroimaging studies of recovery.

**Study aims and hypotheses**

In the current study, we first tested what we consider to be the primary prediction of current models of language recovery after stroke using a relatively large (n=43) sample of patients with chronic (> 1 year) post-stroke aphasia and a sample (n=43) of demographically matched healthy controls. Specifically, we tested the prediction that long-term language outcomes depend on the preservation/restoration of language task-driven activation in canonical language networks. To test this prediction, we first defined the canonical semantic network (CSN) as the set of regions that were more strongly activated by semantic decisions relative to tone decisions in the healthy controls. This bilateral but predominantly left-lateralized network is well-suited for testing the prediction that residual language abilities depend on the preservation/recovery of activity within distributed canonical language networks as it is (1) well-characterized in healthy individuals, and (2) consists of distributed brain regions that include the left angular gyrus (AG)/posterior inferior parietal lobule (pIPL), left middle temporal gyrus (MTG), left parahippocampal gyrus (PHG), left dorsal superior frontal gyrus (dSFG), left inferior frontal gyrus (IFG), and left posterior cingulate cortex (PCC) [Binder et al., 1997; Binder et al., 1999; Binder et al., 2008; Binder et al., 2009; Kim et al., 2011; Szaflarski et al., 2002; Xu et al., 2016]. Thus, we tested whether task-evoked activity in the CSN identified in controls predicted in-scanner performance on the semantic decision task, out-of-scanner performance on verbal fluency tasks, and out-of-scanner performance on a picture naming task in the patient group (with and without controls for lesion volume effects). We note that while the in-scanner and out-of-scanner tasks differ in their emphases on speech comprehension and speech production, respectively, previous studies suggest that activity evoked by speech comprehension tasks may also relate performance on other measures of language function that include speech production tasks [e.g. Saur et al., 2006; Szaflarski et al., 2013]. Other previous studies also suggest that speech production networks (i.e. regions activated during word/verb generation) are also activated (albeit somewhat less strongly) during sentence comprehension tasks [Piervincenzi et al., 2016], and that estimates of language lateralization based on activations measured during both speech production and auditory semantic decision tasks show similar relationships to estimates based on intracarotid amobarbital measurements [Szaflarski et al., 2008].

Additionally, we tested the explanation that strong right hemispheric activation in chronic patients reflects compensation driven by a reduced capacity of the left hemisphere to perform language-related processing. As noted in the previous paragraph, the CSN has bilateral components [e.g. Binder et al., 2009], and thus activation of the right hemispheric network might be expected to increase to compensate for reduced left hemispheric network function [e.g. Hartwigsen et al., 2013]. While previous studies have investigated how activity in the right hemisphere relates to the effects of damage to specific left hemispheric regions [Blank et al., 2003; Sims et al., 2016; Turkeltaub et al.,

2011], we are not aware of any studies that have directly measured the how activation in the right hemisphere relates to the extent of damage sustained by the left hemisphere.

Further, the notion that language-task evoked activation in the right hemisphere is most pronounced in patients with extensive left hemispheric lesions is commonly discussed in the literature [Hamilton et al., 2011; Heiss and Thiel, 2006]. Despite its intuitive appeal, there appears to be little empirical evidence to support this claim. For example, a recent review of the mechanisms of language recovery after stroke [Hamilton et al., 2011] cites only two reports to support the unambiguous assertion that "larger lesions involving eloquent cortex of the left hemisphere are associated with greater recruitment of the right hemisphere during language tasks". One is an earlier review [Heiss and Thiel, 2006] that discusses this idea within the context of a larger hierarchical model of recovery but does not itself provide empirical support for the veracity of this claim in stroke patients, and the other [Kertesz et al., 1979] is a computerized tomography study that shows more extensive left hemispheric damage to be predictive of more severe aphasia and poorer language recovery after stroke, but that does not relate lesion size to any measurement of functional activation. Therefore, we sought to empirically test the hypotheses that right hemispheric language task activation in chronic stroke patients reflects compensation driven by reduced left hemispheric function, and that this effect is most pronounced in patients with the most extensive left hemispheric damage. We expected, based on the literature discussed above, that higher levels of right CSN activation during semantic decisions would be associated with (1) lower levels of left CSN activation, and (2) more extensive left hemispheric damage. As noted above, the CSN is predominantly left-lateralized (i.e. the right and left CSN are not mirror-symmetric), and so we also assessed these effects in right hemispheric regions mirroring the left CSN.

Finally, because we are not aware of any studies that have directly measured the effect of left hemispheric lesion volume on regional activation in the right hemisphere, we characterized regions in the right hemisphere where task-driven responses correlated with left hemispheric lesion volume. To address the question of whether activity in these regions might reflect beneficial compensation that is most pronounced for patients with extensive left hemispheric damage, we further assessed whether relationships between activity in these regions and language measures depended on the extent of left hemispheric damage.

## Materials and Methods

### Participants

All study procedures were approved by the Institutional Review Boards of the participating institutions and were performed in accordance with Declaration of Helsinki ethics principles and principles of informed consent. For the current study, we used functional MRI data collected from 43 chronic post-stroke aphasia patients and 43 healthy controls. Imaging and behavioral data for the post-stroke aphasia patients were collected as part of several separate studies performed by our laboratory. Prior to inclusion, all participants were screened to exclude individuals that had diagnoses of degenerative/metabolic disorders, had severe depression or other psychiatric disorders, were pregnant, were not fluent in English, or had any contraindication to MRI/fMRI. Patients were included in the current study if they had a single left hemispheric stroke resulting in aphasia at least 1 year prior to data collection. Data for the control participants were selected from a database of 150 healthy individuals collected as part of several studies performed by our laboratory, and were selected based on age group (19-29, 30-39, 40-49, 50-59, 60+), handedness as determined by the Edinburgh Handedness Inventory [Oldfield, 1971], and sex to minimize differences in demographics with respect to the patient group. Participant demographics are shown in Table 1. A more detailed characterization of patient demographics is provided in the Supplementary Material (S2).

### Language measures

Prior to MRI scanning, all participants were administered a battery of neuropsychological language assessments. All participants performed the Boston Naming Test (BNT) [Kaplan et al., 2001], Semantic Fluency Test (SFT) [Kozora and Cullum, 1995], and Controlled Oral Word Association Test (COWAT) [Lezak et al., 1995]. The BNT requires patients to name a series of black and white line drawings that contain both animate and inanimate items that vary in frequency of use (e.g. bed vs. abacus), and the number of correctly named pictures serves as a measure of naming ability. The SFT requires patients to generate as many words as they can think of that fit a given category prompt (animals/fruits and vegetables/things that are hot) within a one-minute time limit, and the number of words generated serves as a measure of category fluency. The COWAT requires patients to generate as many words as they can think of that begin with a particular letter (C/F/L) within a one-minute time limit, and the number of words generated serves as a measure of phonemic fluency. Performance scores for the COWAT and SFT were very strongly correlated across patients ($r=0.92$), so they were averaged together to define a single combined measure of verbal fluency. Performance scores for the verbal fluency measure were correlated, albeit less strongly, with performance scores for the BNT ($r=0.76$). Individual patient language task data are provided in the Supplementary Material (S2).

**Neuroimaging data collection**

MRI data were collected at the University of Alabama at Birmingham using a 3T head-only Siemens Magnetom Allegra scanner located in the Civitan International Research Center Functional Imaging Laboratory. These data consisted of 3D high-resolution T1-weighted anatomical scans (TR/TE = 2.3 s/2.17 ms, FOV = 25.6×25.6×19.2 cm, matrix = 256x256, flip angle = 9 degrees, slice thickness = 1mm), and two T2*-weighted gradient-echo EPI pulse functional scans (TR/TE = 2.0 s/38.0 ms, FOV = 24.0x13.6x24.0, matrix = 64x64, flip angle = 70 degrees, slice thickness = 4 mm, 165 volumes per scan). MRI data were also collected at the Cincinnati Children's Hospital Medical Center on a 3T research-dedicated Phillips MRI system located in the Imaging Research Center. These data consisted of 3D high-resolution T1-weighted anatomical scans (TR/TE = 8.1 s/2.17 ms, FOV = 25.0×21.0×18.0 cm, matrix = 252x211, flip angle = 8 degrees, slice thickness = 1mm) and two T2*-weighted gradient-echo EPI pulse sequence functional scans (TR/TE = 2.0 s/38.0 ms, FOV = 24.0x13.6x24.0, matrix = 64x64, flip angle = 70 degrees, slice thickness = 4 mm, 165 volumes per scan).

Functional MRI scans were acquired while participants completed alternating 30 second blocks of a semantic decision/tone decision task. This fMRI task was selected because it has been previously shown to result in robust activation in canonical areas involved in semantic language processing [Binder et al., 1997; Kim et al., 2011], and has been used extensively to evaluate language network activation in healthy and diseased populations that include patients with chronic post-stroke aphasia [e.g. Eaton et al., 2008; Szaflarski et al., 2008; Szaflarski et al., 2011].

The active condition (semantic decision -- SD) was performed 5 times during each scan. Each block of the active condition consisted of 8 trials where participants were presented with spoken English animal names. On each trial, participants decided if the presented animal met the criteria "native to the United States" and "commonly used by humans". If both criteria were satisfied, then the participants responded "1" by using their non-dominant hand to press a button. If both criteria were not satisfied, then the participants responded "2" by using their non-dominant hand to press a second button. The control condition was performed 6 times during each scan. Each block of the control condition (tone decision – TD) consisted of 8 trials where participants were presented with brief sequences of four to seven 500- and 750-Hz tones. On each trial, participants decided if the sequence contained two 750-Hz tones. If the sequence contained two 750 Hz tones, then they pressed the button designated "1" with their non-dominant hand. Otherwise, they pressed the button designated "2" with their non-dominant hand. Each scan lasted 7 minutes and 15 seconds. Prior to completing the task in the MRI scanner, all participants confirmed their understanding of the task by performing a mock run that included a sequence of five sets of tones followed by a sequence of five nouns designating different animals. In-scanner task data were not collected for 4 patients due to hardware issues, and they were excluded from analyses investigating relationships between fMRI activation and in-scanner performance.

**Neuroimaging data processing**

All MRI data were processed using Statistical Parametric Mapping (SPM) [Friston et al., 1995] version 12 running in MATLAB r2014b (The MathWorks, Natick MA, USA). Functional MRI scans were pre-processed using a standard pre-processing pipeline consisting of slice-time correction, realignment/reslicing, co-registration of the fMRI data to the corresponding anatomical scan, unified segmentation with optimized tissue priors for lesioned brains and normalization of the anatomical scan to MNI space

[Ripollés et al., 2012; Seghier et al., 2008], normalization of the functional scan to MNI space using the transformation applied to the anatomical scan, and spatial smoothing using an 8mm full-width half maximum Gaussian kernel. This pipeline enables accurate template registration and normalization even for patients with structural abnormalities such as those observed in stroke patients [Ripollés et al., 2012]. To reduce the potential for motion-related artifacts, functional scans were motion-corrected by replacing volumes with >0.5mm motion with an interpolated volume created from adjacent volumes [Mazaika et al., 2005].

Lesion probability maps were created for each patient using a voxel-based naïve Bayes classification algorithm that was developed by our laboratory and is implemented in the *lesion_gnb* toolbox for SPM12 [Griffis et al., 2016a]. While automated classification with this method compares favorably to manual lesion delineation for large and small lesions [Griffis et al., 2016a], we opted to manually threshold the resulting posterior probability maps to ensure that they precisely reflected the extent of the lesion. Our decision to manually threshold the resulting posterior probability maps was primarily motivated by the potential for automated methods to introduce false positive voxel clusters in patients with small lesions [e.g. Griffis et al., 2016a; Wilke et al., 2011], and while an arbitrary cluster threshold (i.e. 100 voxels) is often sufficient to remove these clusters, it is not guaranteed. Further, as we note in our report describing the validation of our method [Griffis et al., 2016a] it is important to inspect the lesion masks as a quality control step, and by manually thresholding the probability maps produced by the automated classification procedure, we were able to ensure that the final lesion masks accurately reflected each patient's lesion. The resulting binary lesion masks were used to estimate lesion volume and lesion-ROI overlaps, and were used in all further lesion analyses. Lesion frequencies across all 43 patients are shown in Figure 2A. Individual patient lesion images are provided in the Supplementary Material (S2).

For each participant, the pre-processed fMRI data were fit to a general linear model (GLM) where task blocks were modeled as boxcar regressors convolved with a canonical hemodynamic response function (HRF). In order to account for variability in the time-to-peak of the HRF, time and dispersion derivatives were included as basis functions [Meinzer et al., 2013]. Note that since data were collected using a blocked design, separate modeling of correct vs. incorrect trials was not possible. Linear contrast estimate maps were then computed to quantify the difference in activation magnitudes between the active and control conditions. These contrast maps were used for all further functional MRI analyses.

**Statistical Analyses**

Group differences on behavioral measures were assessed for descriptive purposes using two-tailed independent samples t-tests with degrees of freedom adjustment for unequal variances.

To identify regions that showed significant group-level activation for each group, the first-level contrast estimate maps from all individuals in each group were entered into separate second-level $t$ contrasts quantifying the difference in peak HRF magnitude between the SD and TD conditions. Regional activation was considered significant if it survived a combined voxel-level intensity ($p$-value) threshold of 0.01 and a cluster-level extent threshold of $p<0.05$ corrected ($k_{crit}$=99 voxels) to control the whole-brain family-wise error rate (FWE) at 0.05 as determined by 1000 Monte Carlo simulations [Slotnick et al., 2003]. The thresholded map obtained for the control group was then binarized, creating a CSN region-of-interest (ROI) mask where statistically significant voxels had a value of 1 and all other voxels had a value of 0. For each patient, the mean contrast estimate across all voxels within this CSN ROI was then extracted from their first-level contrast estimate map, quantifying the average magnitude of the task-driven response in the CSN. These estimates were used in subsequent ROI correlation analyses. Group differences in activation within the CSN were assessed for descriptive purposes using an independent samples t-test with degrees of freedom adjustment for unequal variances. Lesion-ROI overlaps were calculated for each patient using the lesion masks created during pre-processing. The correlation between total lesion volumes and lesion-ROI overlaps indicated that lesion-ROI overlap was a close linear transformation of lesion volume ($r$=0.93).

To assess the unique relationships between activity in the CSN and language function in chronic patients, the mean contrast estimate quantifying activation within the HC-defined CSN was then entered as an independent variable in 3 separate partial correlation analyses. The dependent variables for each analysis were the performance scores for the in-scanner SD task (% correct responses) and performance scores for the out-of-scanner naming and fluency tasks. Each model included total lesion volume as a covariate to account for variance attributable to lesion volume effects. Analyses were repeated without lesion volume control. Linear correlation analyses assessing imaging-behavior relationships were also performed for the control group using the CSN ROI. To fully characterize relationships between regional language task activation and performance on the in-scanner and out-of-scanner language tasks, three whole-brain multiple regressions (one for each language task) were performed that each included lesion volume as a covariate. Because patient data were collected on different scanners, all correlational analyses for the patient group were repeated with scanner included as a covariate, with and without lesion volume control. These analyses are provided in the Supplementary Material (Supplementary Figure 4).

To test whether the magnitude of activation in right hemispheric portions of the CSN was increased in patients with lower levels of activation in left hemispheric portions of the network, the CSN ROI mask was split into left and right hemispheric components. For each patient, the mean contrast estimate was then extracted from each hemispheric ROI as described above. Linear correlations were assessed between the mean contrast estimates obtained from each hemisphere with and without lesion volume control. Correlations were also assessed between left hemispheric and right hemispheric activation in controls. To explicitly test for this effect in right hemispheric homologues of the left hemisphere network and control for differences in extent between left and right hemispheric portions of the CSN, these analyses were also repeated using a right hemispheric ROI that was defined by simply mirroring the left-hemispheric ROI to the right hemisphere. To assess whether out-of-network right hemispheric homologues of the left hemispheric CSN (i.e. regions that were activated in the left, but not right hemisphere in controls) differed from the right hemispheric CSN with regard to their relationship to left hemispheric CSN activation, a right hemispheric out-of-network homologue ROI was created by subtracting the right hemispheric CSN from the mirrored left hemispheric CSN (see Figure 4A for illustrations of each ROI). Differences in activation between left and right hemispheric networks were assessed with dependent samples t-tests for each group. Between-group differences in activation for left and right hemispheric networks (and differences in activation between left and right hemispheric networks) were assessed with unequal variance t-tests. Between-group differences in the strengths of correlations between left and right CSN ROIs were assessed using *z*-tests [Fisher, 1921].

Lastly, to assess the effect of lesion size on task-driven activation in the right hemisphere, we performed an additional whole-brain linear regression analysis that was restricted to the right hemisphere and included only total lesion volume as a predictor. Post-hoc moderation analyses were performed for right hemisphere clusters that showed significant positive correlations between activation magnitudes and lesion volume in order to assess whether activity in these regions reflected beneficial compensation in patients with larger lesions. For each identified cluster, three separate multiple regression analyses (one for each language measure) were performed that each included mean activation magnitudes for that cluster, total lesion volume (entered as a proportion of the maximum lesion volume), and the interaction term between mean cluster activation and lesion volume. For these post-hoc analyses, only the interaction effects were of interest, since we our hypothesis was that the benefit of recruiting these regions would depend on the extent of left hemispheric damage.

Statistical tests were two-tailed and significance thresholds were set as follows. Results for group comparisons and for each set of ROI-based analyses were considered significant if they survived a Bonferroni-Holm step-down correction procedure to control the FWE at 0.05 across each set of tests, since this method provides good protection against Type I errors and better power than standard Bonferroni correction [Aickin and Gensler, 1996]. T-contrast maps for the effects of interest from each of the whole-brain multiple regression analysis were thresholded at $p<0.01$, uncorrected at the voxel level and multiple comparisons corrected at the cluster level to control the cluster-wise FWE at 0.01 ($k_{crit} = 126$ voxels, as determined by 1000 Monte Carlo simulations) for each contrast. Since power is often lower for analyses of moderation [Jaccard et al., 1990], the *post-hoc* moderation analyses were multiple comparisons corrected using a more lenient False Discovery Rate (FDR) correction threshold of 0.10 [Benjamini and Hochberg, 1995] computed across the interaction term *p*-values for all moderation models.

We note that a recent report [Eklund et al., 2016] has called into question the validity of cluster-based correction methods in fMRI research, particularly when they are used in conjunction with voxel-level thresholds above 0.001. Our choice to use the voxel-level threshold of 0.01 was primarily motivated by its prevalence in the neuroimaging literature of post-stroke aphasia, and because detection power is reduced for group analyses of highly heterogeneous groups such as stroke patients [e.g. Meinzer et al., 2013]. In addition, we note that for the whole-brain regression analyses, we employed a more stringent whole-brain cluster-correction threshold of 0.01 (as opposed to the conventional 0.05 threshold), even when our analyses were restricted to only the right hemisphere. Thus, we have attempted to balance providing control over false positives while maintaining sufficient power to detect effects of interest. In addition, we provide voxel-wise FDR thresholded activation maps and cluster/peak statistics for each group in Supplementary Figure 4 to (1) demonstrate that similar activation patterns are detected at stringent voxel-wise correction thresholds, and (2) provide more precise anatomical information about peak activation locations for each group.

## Results

### Comparisons of overall CSN activation and task performance between patients and controls

Patients performed more poorly on language tasks than controls (in-scanner SD % correct: $t_{74.77}=-4.71$, $p<0.001$, corrected; BNT: $t_{43.38}=-6.80$, $p<0.001$, corrected; fluency: $t_{82.28}=-13.44$, $p<0.001$, corrected). Summary statistics are shown in Figure 2B. The semantic networks recruited by each group are shown in Figure 2C. Cluster peak co-ordinates and statistics for regions activated by each group are provided in Table 2. Controls showed significantly higher levels of overall activation in the CSN than patients ($t_{67.72}=3.98$, $p<0.001$, corrected).

### Region of interest behavioral correlations

Total lesion volume showed negative relationships with each of the language measures (in-scanner SD correct: $r=-0.22$, $p=0.19$, corrected; fluency: $r=-0.58$, $p<0.001$, corrected; naming: $r=-0.49$, $p=0.004$, corrected), although the relationship was not significant for the in-scanner SD task. Lesion volume showed a trend-level but non-significant negative relationship with mean activation in the CSN ($r=-0.27$, $p=0.15$, corrected). The partial correlation analyses revealed significant positive relationships between activity within the CSN and in-scanner performance on the SD task (partial $r=0.47$, $p=0.008$, corrected), out-of-scanner performance on the fluency measure (partial $r=0.62$, $p<0.001$, corrected), and out-of-scanner performance on the naming measure (partial $r=0.49$, $p=0.005$, corrected) that were linearly independent of the effects of total lesion volume (Figure 2D). Relationships between CSN activation and in-scanner performance on the SD task ($r=0.51$, $p=0.005$, corrected), out-of-scanner performance on the fluency measure ($r=0.65$, $p<0.001$, corrected), and out-of-scanner performance on the naming measure ($r=0.55$, $p=0.001$, corrected) were stronger without lesion volume control. These results are shown in Figure 2D.

Additional analyses assessed relationships between mean CSN activation magnitudes and behavioral measures in controls. Mean CSN activation magnitudes showed a significant positive relationship to performance on the in-scanner SD task ($r=0.37$, $p=0.03$, corrected), but did not show significant relationships with performance on the fluency ($r=0.09$, $p=1.0$, corrected) or naming ($r=0.09$, $p=1.0$, corrected) tasks (Figure 2D).

**Whole-brain behavioral correlations**

The results for the whole-brain multiple regressions for each language measure that included total lesion volume as a covariate are shown in Figure 3A. Cluster peak locations and statistics for each whole-brain regression are provided in Table 3. Figure 3B shows regions that showed positive relationships between activation and performance on all three language measures, and Figure 3C shows regions that showed positive relationships between activation and any of the language measures relative to the locations of regions associated with the CSN identified in healthy controls. Whole-brain analyses repeated without including lesion volume as a covariate showed very similar results (Supplemental Tables 7-9).

**Correlations between left and right hemispheric activation magnitudes**

Partial correlation analyses (controlling for lesion volume) for the patient group revealed a strong positive relationship between activation magnitudes in the left CSN and activation magnitudes in the right CSN (partial $r=0.83$, $p<0.001$, corrected). Nearly identical results were obtained without lesion volume control ($r=0.83$, $p<0.001$, corrected). A strong positive relationship was also observed between activation magnitudes in the left CSN and activation magnitudes in the mirrored left CSN in the right hemisphere with (partial $r=0.81$, p<0.001, corrected) and without ($r=0.80$, $p<0.001$, corrected) lesion volume control. However, correlations were much weaker between activation magnitudes in the left CSN and activation magnitudes in out-of-network homologues of the left hemispheric CSN in the right hemisphere with (partial $r=0.66$, $p<0.001$, corrected) and without ($r=0.65$, $p<0.001$, corrected) lesion volume control. Left and right hemispheric ROIs are shown in Figure 4A. Scatterplots illustrating the relationships between residual left and right hemispheric activation magnitudes (i.e. after partialling out lesion volume) are shown in Figure 4B (top row).

Correlation analyses for the control group also revealed a strong positive relationship between activation magnitudes in the left CSN and activation magnitudes in the right CSN ($r=0.75$, $p<0.001$, corrected). A weaker relationship was observed between activation magnitudes in the left CSN and activation magnitudes in the mirrored left CSN in the right hemisphere ($r=0.43$, $p=0.008$, corrected). No significant relationship was observed between activation magnitudes in the left CSN and activation magnitudes in out-of-network homologues of the left CSN in the right hemisphere ($r=-0.04$, $p=0.79$, corrected). Scatterplots illustrating these relationships are shown in Figure 4B.

The strengths of correlations between left CSN and right CSN activation magnitudes did not significantly differ between patients and controls ($z=0.52$, $p=0.6$, corrected), but the strengths of correlations between activation magnitudes in the left CSN and mirrored left CSN in the right hemisphere ($z=2.98$, $p=0.006$, corrected) and between activation magnitudes in the left CSN and out-of-network homologues of the left

CSN in the right hemisphere ($z=3.72, p<0.001$, corrected) were significantly stronger in patients than in controls.

**Comparisons of left and right hemispheric activation magnitudes**

Activation magnitudes for each group are shown in Figure 4C. Neither patients ($t_{42}=0.96, p=0.34$, corrected) nor controls ($t_{42}=1.72, p=0.19$, corrected) showed significant differences in activation magnitudes between the left CSN and right CSN. Both patients ($t_{42}=3.89, p=0.001$, corrected) and controls ($t_{42}=9.22, p<0.001$, corrected) showed significantly larger activation magnitudes in the left CSN compared to the mirrored left CSN in the right hemisphere. Both patients ($t_{42}=5.42, p<0.001$, corrected) and controls ($t_{42}=12.92, p<0.001$, corrected) also showed significantly larger activation magnitudes in the right CSN compared out-of-network homologues of the left CSN in the right hemisphere.

Patients had significantly lower activation magnitudes in both the left ($t_{66.37}=-3.60, p<0.001$, corrected) and right ($t_{74.28}=-4.23, p<0.001$, corrected) CSN than controls. Patients also had significantly lower activation magnitudes in the mirrored left CSN in the right hemisphere ($t_{67.17}=-2.92, p=0.01$, corrected). Patient and control activation magnitudes in out-of-network homologues of the left CSN in the right hemisphere did not significantly differ ($t_{60.15}=0.47, p=1.0$, corrected).

The difference between activation in the left CSN and the right CSN was not significantly larger in controls than in patients ($t_{81.89}=0.38, p=1.0$, corrected). Controls showed a significantly larger difference in activation between the left hemispheric CSN and the mirrored left CSN in the right hemisphere ($t_{83.82}=3.99, p<0.001$, corrected). Controls also showed a significantly larger difference between activation magnitudes in the right CSN and out-of-network homologues of the left CSN in the right hemisphere ($t_{82.15}=6.26, p<0.001$, corrected).

**Lesion volume effects on activation magnitudes**

Activation magnitudes in both the left CSN ($r=-0.24, p=0.35$, corrected) and right CSN ($r=-0.31, p=0.18$, corrected) showed non-significant negative relationships to lesion volume. A weaker negative relationship was observed between lesion volume and activation magnitudes in the mirrored left CSN in the right hemisphere (r=-0.12, p=0.90, corrected). Lesion volume effects showed a near-zero correlation with activation magnitudes in out-of-network homologues of the left CSN in the right hemisphere (r=-0.01, p=0.97, corrected). These results are shown in Figure 4D.

The results from the whole-brain regression analysis assessing the effects of lesion volume on activation in the right hemisphere are shown in Figure 5A. Cluster statistics and peak locations are displayed in Table 4. All of the regions showing positive relationships with lesion volume fell outside of the canonical semantic network in right frontal cortex (Figure 5B). Notably, with the exception of the right middle frontal gyrus (MFG), these regions did not correspond to out-of-network homologues of the left CSN. In contrast, several of the regions showing negative relationships between activation and lesion volume overlapped with the CSN (Figure 5B).

To assess whether there was a benefit of activating out-of-network right-hemisphere regions that depended on the extent of left hemispheric damage, we performed additional *post-hoc* regression analyses to assess whether the lesion size might moderate the effect of activity in each right hemispheric cluster showing positive correlations to total lesion volume (hot clusters in Figure 5A). This revealed significant interactions between lesion volume and right SMA activation for the fluency ($t_{39}$=2.97, FDRp = 0.04) and naming ($t_{39}$=2.82, FDRp = 0.04) measures (Figure 5C), and between lesion volume and right IFG pars opercularis activation for the fluency ($t_{39}$=2.50, FDRp = 0.07) measure. For each of the observed interaction effects, higher levels of activation in the right hemispheric cluster was associated with lower language test scores for patients with smaller lesions, but with higher language test scores for patients with larger lesions (Figure 5C). Note that the high and low values shown in Figure 5C are calculated for +1 and -1 standard deviations from the mean of each variable.

## Discussion

Functional neuroimaging studies of post-stroke aphasia are foundational to theories of language recovery after stroke, but this theoretical foundation is primarily composed of small studies. To address this shortcoming and assess the validity of predictions based on current theories of recovery, we tested what we consider to be their primary prediction – that successful long-term language recovery depends on the preservation and/or restoration of language processing in canonical language networks. Importantly, we tested this prediction in one of the largest samples of post-stroke aphasia patients studied to date in the functional neuroimaging literature, and utilized statistical controls to reduce the potential for confounding effects related to lesion volume on imaging-behavior relationships. We also attempted to bring clarity to contested issues in the literature – how language task-driven activity in the unaffected right hemisphere relates to left hemispheric damage and function, and whether the recruitment of right hemispheric regions by patients with extensive left hemispheric damage supports residual language function. We discuss our results and their relation to the broader literature in the following sections.

**Canonical language network contributions to post-stroke language function**

Drawing from the broader functional neuroimaging literature of language recovery after stroke, we expected that activation of the CSN for language processing would positively predict language functions in chronic patients. The results of both our ROI-based and whole-brain analyses matched this expectation, and indicate that activation of the CSN is a moderate-to-strong predictor of language function in chronic patients (Figures 2 and 3). While activation magnitudes in this network only predicted performance on the in-scanner task for controls, they predicted performance on all three language measures for patients (Figure 2D). Further, our whole-brain analyses of the patient data revealed that the positive relationships between regional task-driven activation magnitudes and language task performance were primarily localized within or adjacent to the CSN identified in controls (Figure 3). Thus, our results corroborate the general implications of previous functional neuroimaging studies of language outcomes in patients with LMCA stroke [Fridriksson et al., 2010; Fridriksson et al., 2012; Heiss et al., 1999; Karbe et al., 1998; van Oers et al., 2010; Rosen et al., 2000; Saur et al., 2006; Szaflarski et al., 2013]. Importantly, this result supports the emphasis of current theory on the preservation and/or restoration of function in canonical language networks as a key factor that enables successful long-term language recovery after stroke.

Notably, our results indicate that despite the fact that the auditory semantic decision task lacks an expressive language component, the level of CSN activation evoked was a good predictor of expressive language capacity in our patient sample (Figure 2D/Figure 3). As noted in the introduction, previous studies suggest that relationships between activation evoked by auditory (i.e. receptive) language tasks and language function in chronic stroke patients may not be specific to receptive language functions [e.g. Saur et al., 2006; Szaflarski et al., 2013]. Our results suggest that activation in the CSN evoked by auditory semantic decisions may provide an index of general language network preservation/recovery.

This interpretation is consistent with evidence that the abrupt and catastrophic disruption of neural function by stroke leads to widespread disruptions of communication and regulation in distributed brain networks [Baldassarre et al., 2016; Geranmayeh et al., 2016; He et al., 2007; Ovadia-Caro et al., 2013; Siegel et al., 2016], and that the preservation/restoration of typical function in canonical language networks is indicative of successful language recovery [Heiss et al., 1999; Saur et al., 2006]. Indeed, post-stroke deficits in complex cognitive functions [Siegel et al., 2016] that include language [Geranmayeh et al., 2016] and attention [Baldassarre et al., 2016; He et al., 2007] may be conceptualized as behavioral manifestations of dysfunction in large-scale functional brain networks. Along these lines, optimal functional recovery after stroke may depend on the preservation/restoration of functional dynamics that most strongly resembles those observed in the pre-stroke brain [Carter et al., 2012].

Thus, the magnitude of task-evoked responses in the distributed CSN may index the degree to which pre-stroke functional dynamics are preserved/restored in patients with chronic post-stroke aphasia. While discussions of how distributed networks contribute to recovery from stroke often focus on measures of intrinsic (i.e. resting state) network function [e.g. Carter et al., 2012], there is substantial evidence that task-evoked brain networks are strongly influenced by the intrinsic network architecture [Binder et al., 1999; Binder, 2012; Cole et al., 2014; Dosenbach et al., 2006; Dosenbach et al., 2007; Fox et al., 2005; Fox et al., 2006; Muhle-Karbe et al., 2015; Xu et al., 2016], and thus it is possible that the presence of strong activation in the CSN during auditory semantic decisions may be indicative of well-preserved and/or successfully restored intrinsic network dynamics. We must emphasize, however, that this explanation is tentative and must be confirmed by future studies that are capable of relating task-evoked responses in canonical language networks to intrinsic network dynamics in patients with chronic post-stroke aphasia.

**Co-activation of left and right hemispheric networks and right hemispheric effects of lesion volume**

Based on current models of language recovery after stroke [Hamilton et al., 2011; Heiss and Thiel, 2006; Saur et al., 2006; Saur and Hartwigsen, 2012], we expected that patients with lower levels of activation in the left CSN would show higher levels of activation in the right CSN. Assuming that this would reflect compensatory up-regulation of the right CSN following extensive left hemispheric damage, we also expected that activation in the right CSN would be increased in patients with larger left hemispheric lesion volumes. However, our results did not bear out this prediction. Rather, we found that for both patients and controls, activation magnitudes in the right CSN were strongly positively correlated with activation magnitudes in the left CSN (Figure 4B). Further, in patients, activation magnitudes in both the left and right CSNs showed negative but non-significant correlations with lesion volume (Figure 4C). This suggests that the right CSN and left CSN form a coherent functional unit that is activated by semantic processing. Indeed, a basic role of certain right hemispheric regions in language processing is suggested by findings that right hemispheric activation during semantic decision [Donnelly et al., 2011], semantic comprehension [van Ettinger-Veenstra et al., 2010], sentence completion [van Ettinger-Veenstra et al., 2012], and word fluency tasks [van Ettinger-Veenstra et al., 2012] correlates with various measures of language function in healthy individuals.

Notably, a recent ALE meta-analysis of 12 functional neuroimaging studies of aphasic patients (total n=105) and healthy controls (total n=129) suggests that specific regions in both hemispheres are consistently recruited by aphasic patients across different language task paradigms [Turkeltaub et al., 2011]. Turkeltaub and colleagues [2011] thus proposed that activity in most right hemispheric regions, with the notable exception of the IFG pars triangularis, likely reflects either compensatory recruitment or co-activation with homotopic areas in the left hemisphere. Our finding that activation magnitudes in the left and right CSN were strongly correlated suggests that at least with regard to the CSN, activity in right hemispheric portions of canonical networks likely reflects co-activation rather than compensation for left hemispheric damage. The finding that neither patients nor controls showed significant differences between the left and right CSN supports this conclusion.

However, we also found that both patients and controls showed significantly higher activation in the left CSN than in homologous areas in the right hemisphere, and this effect was most pronounced for homologous areas that were not associated with the right CSN (Figure 5C). Further, both patients and controls showed significantly higher levels of activity in the right CSN compared to these out-of-network left CSN homologues (Figure 5C). This suggests that co-activations are spatially constrained to the subset of right hemispheric regions that typically activate as part of the CSN. While the correlation between activation magnitudes in the left CSN and out-of-network right hemispheric homologues was substantially weaker than the correlation between activation magnitudes in the left CSN and right CSN in patients (Figure 5B), controls showed essentially no relationship between activation magnitudes in the left CSN and out-of-network right hemispheric homologues (Figure 5B). This suggests that inter-hemispheric co-activations may be less spatially constrained in stroke patients than in healthy controls. Speculatively, this may reflect weakened inter-hemispheric inhibition, as this has been proposed as a potential source of increased activations in right hemispheric homologues of left hemispheric language areas in patients with post-stroke aphasia [Hamilton et al., 2011; Heiss and Thiel, 2006].

Nonetheless, our data support a functional distinction between right hemispheric regions that show robust co-activation with distributed language networks and out-of-network right hemispheric homologues of left hemispheric language areas, although conclusions regarding the source of this distinction cannot be drawn from this study. Future studies using task-based (i.e. effective) connectivity metrics to assess how activation in these regions relate to interactions between left and right hemispheric networks are necessary to address such questions.

**Right hemispheric compensation in patients with extensive left hemispheric damage**

Recent evidence from structural MRI studies also suggests a compensatory role of certain right hemispheric regions in supporting language function in chronic patients. For example, grey matter volume in right dorsal stream temporo-parietal areas is increased in chronic post-stroke aphasia patients relative to both healthy controls and chronic left hemispheric stroke patients without aphasia [Xing et al., 2015]. Importantly, higher grey matter volume in these regions is associated with better residual language functions in chronic post-stroke aphasia patients [Xing et al., 2015]. Similarly, increased fractional anisotropy (FA) has been observed in the right IFG pars opercularis for chronic post-stroke aphasia patients relative to healthy controls, and higher FA in this region is associated with better speech production abilities for patients [Pani et al., 2016]. Our findings add to this literature by showing that functional activation in specific right frontal regions during language task performance is associated with better expressive language abilities in chronic post-stroke aphasia patients with extensive left hemispheric damage.

Notably, larger lesions were associated with higher levels of activity in several right frontal areas, including the right inferior frontal gyrus pars opercularis and supplementary motor area, both of which were not part of the CSN (Figure 5A/B). Activation magnitudes in the right IFG pars opercularis and right SMA showed relationships to out-of-scanner language measures that differed in direction between patients with larger vs. smaller lesions (Figure 5C). One explanation as to why stronger activations of these regions were associated with poorer performance for patients with smaller lesions is that activity in these regions interferes with the functions of task-relevant areas when canonical regions are intact, but supports residual language function when canonical regions are damaged. Alternately, stronger activations of these regions for patients with smaller lesions may be reflective of focal damage to critical left hemispheric areas that impede the restoration of canonical network function during recovery, and result in chronic language deficits comparable to those observed in patients with more widespread left hemispheric damage. Based on evidence that both the right IFG and right SMA support language processing during early recovery, but show reduced involvement in later stages when canonical networks are recovered [Saur et al., 2006], we consider this to be most likely. If this is the case, the identification of such critical regions by future studies could provide important insights into the factors contributing to poor long-term language recovery after stroke.

Along these lines, we note that previous studies suggest that activation in the right IFG pars opercularis is increased in patients with lesions affecting the left IFG [Blank et al., 2003; Turkeltaub et al., 2011]. In addition, maintained right frontal activation has been previously reported in patients with left posterior temporal lesions that recovered less successfully than patients with lesions affecting the left frontal cortices or the left basal ganglia [Heiss et al., 1999]. Given that (1) our results indicate that larger lesions correlated with larger activation magnitudes in right frontal cortices, and (2) activation magnitudes in right IFG/SMA were associated with better expressive language functions in patients with larger lesions, we speculate that the observed benefit of activating the right IFG/right SMA in patients with larger lesions may reflect beneficial compensation that supports expressive language functions (e.g. top-down selection and sequencing). Based on the findings of Heiss and colleagues [1999], we further speculate that patients with lesions affecting left posterior temporal regions may also recruit these regions, but that these patients may show reduced benefit from their involvement, presumably resulting from an inability to restore function in other distributed portions of canonical networks. We stress that this must be regarded as speculative, as the current study did not directly investigate relationships between lesion location and right hemispheric activation. Rather, the methods used in this study were primarily intended to address the question of how lesion extent relates to right hemispheric activation. Because the methodology of this study was not ideal for addressing questions about how lesion location relates to regional activation, future studies using methodology better suited to addressing this question are necessary before strong conclusions can be drawn about how the site of damage relates to right hemispheric activations.

**A common network supporting comprehension, naming, and fluency**

Activity in a subset of regions (shown in Figure 3B) correlated positively with performance on all of the language measures, suggesting that this set of regions may support processes critical for language processing after stroke. This set of regions included the left dorsal dSFG, the left AG/superior lateral occipital gyrus, bilateral precuneus and PCC, bilateral PHG and fusiform gyri, and the right temporal pole. These regions are commonly associated with the default mode network [Dosenbach et al., 2007; Fox et al., 2005; Raichle et al., 2001; Raichle and Snyder, 2007; Vincent et al., 2008]. Based on observations that this network is more active during both semantic processing tasks and task-free resting states relative to tasks that do not involve semantic processing (and generally more active during semantic processing than during rest), this network has been proposed to form a core "conceptual network" that is involved in both explicit semantic processing and ongoing manipulations of memory/conceptual representations in the absence of explicit tasks [Binder et al., 1999; Binder et al., 2009; Leech and Sharp, 2014]. Notably, the PCC, precuneus, and AG are functionally connected to the PHG and co-activate with parahippocampal areas during memory retrieval [Sestieri et al., 2011]. The AG [Uddin et al., 2010] and PCC [Greicius et al., 2009] also possess direct structural connections to medial temporal lobe structures such as the hippocampus that play a critical role in memory. Speculatively, since each of the language measures utilized in this study involved a memory component (e.g. recalling facts about animals for the in-scanner semantic decision task, recalling words that begin with a given letter or fit a given category for the COWAT/SFT, and recalling names of visual objects for the BNT), it is possible that activity in these regions reflects their role in supporting memory access and manipulation. In addition, recent intrinsic functional connectivity evidence suggests that regions such as the left dSFG and left AG may function as hubs that support ongoing interactions among canonical language modules (i.e. perisylvian language areas), executive control modules (i.e. fronto-parietal network), and memory/simulation modules (i.e. default-mode network) [Xu et al., 2016].

      Additionally, the right anterior temporal lobe has previously been reported to support auditory verbal comprehension in patients with damage to left posterior temporal areas, and may represent an independent module capable of auditory language processing when left posterior temporal structures are compromised [Crinion and Price, 2005]. When connectivity between left and right anterior temporal areas is preserved, left frontal access to inputs processed by the right anterior temporal lobe may provide a

compensatory mechanism for achieving top-down modulations (e.g. selection or integration) of semantic content via inter-hemispheric pathways [Warren et al., 2009]. In addition, medial posterior default mode regions frequently interact with elements of other functional networks [de Pasquale et al., 2012] and possess diverse structural connections to language-relevant cortical (i.e. inferior parietal, superior temporal, anterior cingulate, and supplementary/pre-motor cortices) and subcortical (including the striatum and multiple thalamic nuclei) areas whose connections may be disrupted by LMCA stroke [Cavanna and Trimble, 2006; Greicius et al., 2009]. Thus, this set of regions could potentially act to relay information among surviving language areas when primary pathways are affected by stroke, providing a potential "back-up" interface among bilateral frontal, temporal, and parietal areas to support manipulations of memory representations and/or speech inputs when canonical cortico-cortical pathways (e.g. the arcuate fasciculus) are no longer viable. While plausible, and consistent with a proposed role of these regions as cross-network "connectors" [Xu et al., 2016], this should be regarded as speculation, as delineating the precise roles these regions play in supporting language processing after stroke is beyond the scope of this study.

**Lesion volume effects in functional neuroimaging of aphasia recovery**

Because we used statistical controls to account for lesion volume effects, our results are not likely to be driven by differences in lesion size or overall damage to the network among patients. Despite the potential for lesion volume differences to introduce bias, we note that our results were nearly identical (albeit with somewhat larger effects) when we did not use controls for lesion volume. This may reflect the relatively low correlation between lesion volume and canonical network activation in this sample. Nonetheless, while we contend that the use of statistical controls is important to reduce the potential for biases related to differences in lesion extent, it is possible that lesion volume is too coarse of a measure to accomplish this goal. An alternative approach may be to utilize more spatially sensitive measures, such as the percent of damage sustained by different anatomical areas [e.g. Xing et al., 2015]. Determining the optimal solution to account for lesion biases in functional neuroimaging studies of stroke patients may ultimately require dedicated studies using simulated data to understand the severity of biases likely to be introduced by varying levels of brain damage and to allow for an objective comparison of different approaches for mitigating such biases.

**Limitations**

This study has several limitations that must be acknowledged. First, as with many functional imaging studies, it is not possible to definitively conclude that activation in any given region is necessary to perform the task. Further, even if this were possible, it would not be possible to conclude from these data which aspect of the task a hypothetical "task-critical" region might support. Thus, while our results are interpreted as indicating that canonical network regions likely support residual language abilities after left hemispheric stroke, and that specific out-of-network regions in the right hemisphere may compensate for extensive left hemispheric damage, conclusions about the nature of the contributions of these regions to task performance cannot be considered definitive. Second, while the current study suggests that extensive left hemispheric damage is associated with the recruitment of select right frontal regions to accomplish semantic processing, it cannot address important questions such as how damage to specific regions in the left hemisphere influences semantic processing or affects out-of-network responses in the right hemisphere; future research is necessary to address these questions. Third, altered vascular dynamics in chronic stroke patients have the potential to lead to changes in neurovascular coupling and influence the BOLD response, and this may lead to reductions in measured activation magnitudes for stroke patients relative to controls [e.g. D'Esposito et al., 2003; Veldsman et al., 2015]. It is necessary to consider the potential for such effects when interpreting the results of this and other studies using functional MRI to study clinical populations such as stroke patients. Fourth, this study was not intended to address the question of how lesion location relates to right hemispheric activation. We stress that strong conclusions regarding this relationship cannot be drawn from the results of this study. Finally, the current study does not allow for conclusions regarding the role of functional interactions among regions associated with the CSN, between left and right hemispheric portions of the CSN, or between the CSN and other networks in supporting language recovery after stroke, as it is limited to assessments of co-activation during task performance. Future studies using functional and/or effective connectivity measures are necessary to enable such conclusions.

**Conclusions**

In conclusion, this study aimed to test current theories of language recovery after stroke and bring clarity to discrepancies in the literature. Our results confirm the prediction that fMRI activation in the canonical semantic network predicts performance on multiple measures of language function in chronic patients independently of lesion volume effects, and reveal a core set of default mode regions that likely play a key role in supporting basic aspects of residual language functions in the years after stroke. Our results also suggest that left and right hemispheric portions of the semantic network co-activate to accomplish language processing after stroke, contrary to the notion that activity in right hemispheric homologues of the left hemispheric network is up-regulated to compensate for damage to the left hemispheric network. In contrast, our results suggest that patients with the most damage to the left hemispheric network recruit specific right hemispheric areas that have been previously shown to play a primary role in supporting early recovery. We suspect that these regions may support top-down selection or motor aspects of residual language function during the chronic stage in these patients. The findings described here emphasize the importance of canonical language networks for supporting long-term language recovery after stroke, clarify the contributions of in- and out- of network right hemispheric areas to language recovery and their relationships to left hemisphere damage, and provide a more stable foundation for future studies of language recovery after stroke.

**Acknowledgements**
Amber Martin
Christi Banks
NIH R01 HD068488
NIH R01 NS048281

# References

Aickin M, Gensler H (1996): Adjusting for multiple testing when reporting research results: The Bonferroni vs Holm methods. Am J Public Health 86:726–728.

Allendorfer J, Kissela B, Holland S, Szaflarski J (2012): Different patterns of language activation in post-stroke aphasia are detected by overt and covert versions of the verb generation fMRI task. Med Sci Monit Monit 18.

Benjamini Y, Hochberg Y (1995): Controlling the False Discovery Rate: A Practical and Powerful Approach to Multiple Testing. J R Stat Soc Ser B 57:289–300.

Baldassarre A, Ramsey L, Rengachary J, Zinn K, Siegel JS, Metcalf N V., Strube MJ, Snyder AZ, Corbetta M, Shulman GL (2016): Dissociated functional connectivity profiles for motor and attention deficits in acute right-hemisphere stroke. Brain 139:2024–2038.

Belin P, Van Eeckhout P, Zilbovicius M, Remy P, François C, Guillaume S, Chain F, Rancurel G, Samson Y (1996): Recovery from nonfluent aphasia after melodic intonation therapy: a PET study. Neurology 47:1504–1511.

Binder JR, Frost JA, Hammeke TA, Cox RW, Rao SM, Prieto T (1997): Human brain language areas identified by functional magnetic resonance imaging. J Neurosci 17:353–362.

Binder JR (2012): Task-induced deactivation and the "resting" state. Neuroimage 62:1086–1091.

Binder JR, Desai RH, Graves WW, Conant LL (2009): Where is the semantic system? A critical review and meta-analysis of 120 functional neuroimaging studies. Cereb Cortex 19:2767–2796.

Binder JR, Swanson SJ, Hammeke TA, Sabsevitz DS (2008): A comparison of five fMRI protocols for mapping speech comprehension systems. Epilepsia 49:1980–1997.

Binder J, Frost J, Hammeke T, Bellgowan PSF, Rao SM, Cox RW (1999): Conceptual processing during the conscious resting state: A functional MRI study. J Cogn Neurosci 11:80–93.

Blank SC, Bird H, Turkheimer F, Wise RJS (2003): Speech production after stroke: The role of the right pars opercularis. Ann Neurol 54:310–320.

Butler RA, Lambon Ralph MA, Woollams AM (2014): Capturing multidimensionality in stroke aphasia: mapping principal behavioural components to neural structures. Brain:3248–3266.

Carter AR, Shulman GL, Corbetta M (2012): Why use a connectivity-based approach to study stroke and recovery of function? Neuroimage 62:2271–2280.

Cavanna AE, Trimble MR (2006): The precuneus: a review of its functional anatomy and behavioural correlates. Brain 129:564–83.

Charidimou A, Kasselimis D, Varkanitsa M, Selai C, Potagas C, Evdokimidis I (2014): Why is it difficult to predict language impairment and outcome in patients with aphasia after stroke? J Clin Neurol 10:75–83.

Cheng B, Forkert ND, Zavaglia M, Hilgetag CC, Golsari A, Siemonsen S, Fiehler J, Pedraza S, Puig J, Cho T-H, Alawneh J, Baron J-C, Ostergaard L, Gerloff C, Thomalla G (2014): Influence of Stroke Infarct Location on Functional Outcome Measured by the Modified Rankin Scale. Stroke 45:1695–1702.

Cole MW, Bassett DS, Power JD, Braver TS, Petersen SE (2014): Intrinsic and task-evoked network architectures of the human brain. Neuron 83:238–251.

D'Esposito M, Deouell LY, Gazzaley A (2003): Alterations in the BOLD fMRI signal with ageing and disease: a challenge for neuroimaging. Nat Rev Neurosci 4:863–872.

Donnelly KM, Allendorfer JB, Szaflarski JP (2011): Right hemispheric participation in semantic decision improves performance. Brain Res 1419:105–116.

Dosenbach NUF, Visscher KM, Palmer ED, Miezin FM, Wenger KK, Kang HC, Burgund ED, Grimes AL, Schlaggar BL, Petersen SE (2006): A Core System for the Implementation of Task Sets. Citeulike:JOUR. Neuron 50:799–812.

Dosenbach NU, Fair DA, Miezin FM, Cohen AL, Wenger KK, Dosenbach RAT, Fox MD, Snyder AZ, Vincent JL, Raichle ME, Schlaggar BL, Petersen SE (2007): Distinct brain networks for adaptive and stable task control in humans. Proc Natl Acad Sci U S A 104:11073–8.

Eaton KP, Szaflarski JP, Altaye M, Ball AL, Kissela BM, Banks C, Holland SK (2008): Reliability of fMRI for studies of language in post-stroke aphasia subjects. Neuroimage 41:311–22.

Eklund A, Nichols TE, Knutsson H (2016): Cluster failure: Why fMRI inferences for spatial extent have inflated false-positive rates. Proc Natl Acad Sci 113:201602413.

van Ettinger-Veenstra HM, Ragnehed M, Hällgren M, Karlsson T, Landtblom A-M, Lundberg P, Engström M (2010): Right-hemispheric brain activation correlates to language performance. Neuroimage 49:3481–8.

van Ettinger-Veenstra H, Ragnehed M, McAllister A, Lundberg P, Engström M (2012): Right-hemispheric cortical contributions to language ability in healthy adults. Brain Lang 120:395–400.

Fisher R (1921): On the probable error of a coefficient of correlation deduced from a small sample. Metron 1:3–32.

Fox MD, Corbetta M, Snyder AZ, Vincent JL, Raichle ME (2006): Spontaneous neuronal activity distinguishes human dorsal and ventral attention systems. Proc Natl Acad Sci U S A 103:10046–51.

Fox MD, Snyder AZ, Vincent JL, Corbetta M, Van Essen DC, Raichle ME (2005): The human brain is intrinsically organized into dynamic, anticorrelated functional networks. Proc Natl Acad Sci U S A 102:9673–8.

Fridriksson J, Bonilha L, Baker JM, Moser D, Rorden C (2010): Activity in preserved left hemisphere regions predicts anomia severity in aphasia. Cereb Cortex 20:1013–9.

Fridriksson J, Richardson JD, Fillmore P, Cai B (2012): Left hemisphere plasticity and aphasia recovery. Neuroimage 60:854–63.

Friston KJ, Holmes AP, Worsley KJ, Poline J-P, Frith CD, Frackowiak RSJ (1995): Statistical parametric maps in functional imaging: A general linear approach. Hum Brain Mapp 2:189–210.

Geranmayeh F, Leech R, Wise RJS (2016): Network dysfunction predicts speech production after left hemisphere stroke. Neurology 86:1296–1305.

Griffis JC, Allendorfer JB, Szaflarski JP (2016a): Voxel-based Gaussian naïve Bayes classification of ischemic stroke lesions in individual T1-weighted MRI scans. J Neurosci Methods 257:97–108.

Griffis JC, Nenert R, Allendorfer JB, Szaflarski JP (2016b): Interhemispheric Plasticity following Intermittent Theta Burst Stimulation in Chronic Poststroke Aphasia. Neural Plast 2016:20–23.

Greicius MD, Supekar K, Menon V, Dougherty RF (2009): Resting-state functional

connectivity reflects structural connectivity in the default mode network. Cereb Cortex 19:72–8.

Hamilton RH, Chrysikou EG, Coslett B (2011): Mechanisms of aphasia recovery after stroke and the role of noninvasive brain stimulation. Brain Lang 118:40–50.

Hartwigsen G, Saur D, Price CJ, Ulmer S, Baumgaertner A, Siebner HR (2013): Perturbation of the left inferior frontal gyrus triggers adaptive plasticity in the right homologous area during speech production. Proc Natl Acad Sci U S A 110:16402–7.

He BJ, Snyder AZ, Vincent JL, Epstein A, Shulman GL, Corbetta M (2007): Breakdown of functional connectivity in frontoparietal networks underlies behavioral deficits in spatial neglect. Neuron 53:905–18.

Heiss WD, Kessler J, Thiel A, Ghaemi M, Karbe H (1999): Differential capacity of left and right hemispheric areas for compensation of poststroke aphasia. Ann Neurol 45:430–438.

Heiss W, Thiel A (2006): A proposed regional hierarchy in recovery of post-stroke aphasia. Brain Lang 98:118–23.

Jaccard J, Wan CK, Turrisi R (1990): The Detection and Interpretation of Interaction Effects Between Continuous Variables in Multiple Regression. Multivariate Behav Res 25:467–478.

Kaplan E, Goodglass H, Weintraub S, Segal O, van Loon-Vervoorn A (2001): Boston naming test. Pro-ed.

Karbe H, Thiel A, Weber-luxenburger G, Kessler J, Heiss W (1998): Brain Plasticity in Poststroke Aphasia : What Is the Contribution of the Right Hemisphere? Brain Lang 230:215–230.

Karnath HO, Berger MF, Kuker W, Rorden C (2004): The anatomy of spatial neglect based on voxelwise statistical analysis: A study of 140 patients. Cereb Cortex 14:1164–1172.

Kertesz A, Harlock W, Coates R (1979): Computer tomographic localization, lesion size, and prognosis in aphasia and nonverbal impairment. Brain Lang 8:34–50.

Kim KK, Karunanayaka P, Privitera MD, Holland SK, Szaflarski JP (2011): Semantic association investigated with functional MRI and independent component analysis. Epilepsy Behav 20:613–622.

Kozora E, Cullum CM (1995): Generative naming in normal aging: Total output and qualitative changes using phonemic and semantic constraints. Clin Neuropsychol 9:313–320.

Kümmerer D, Hartwigsen G, Kellmeyer P, Glauche V, Mader I, Klöppel S, Suchan J, Karnath HO, Weiller C, Saur D (2013): Damage to ventral and dorsal language pathways in acute aphasia. Brain 136:619–629.

Lazar RM, Speizer AE, Festa JR, Krakauer JW, Marshall RS (2008): Variability in language recovery after first-time stroke. J Neurol Neurosurgery, Psychiatry 79:530–534.

Leech R, Sharp DJ (2014): The role of the posterior cingulate cortex in cognition and disease. Brain 137:12–32.

Lezak MD, Howieson DB, Loring DW, Hannay JH, Fischer JS (1995): Neuropsychological assessment (3) Oxford University Press. New York.

Maas MB, Lev MH, Ay H, Singhal AB, Greer DM, Smith WS, Harris GJ, Halpern EF, Koroshetz WJ, Furie KL (2012): The Prognosis for Aphasia in Stroke. J Stroke Cerebrovasc Dis 21:350–357.

Mattioli F, Ambrosi C, Mascaro L, Scarpazza C, Pasquali P, Frugoni M, Magoni M, Biagi L, Gasparotti R (2014): Early aphasia rehabilitation is associated with functional reactivation of the left inferior frontal gyrus: a pilot study. Stroke 45:545–52.

Mazaika PK, Whitfield S, Cooper JC (2005): Detection and repair of transient artifacts in fMRI data. Neuroimage 26:S36.

Meinzer M, Beeson PM, Cappa S, Crinion J, Kiran S, Saur D, Parrish T, Crosson B, Thompson CK (2013): Neuroimaging in aphasia treatment research: consensus and practical guidelines for data analysis. Neuroimage 73:215–24.

Meltzer JA, Wagage S, Ryder J, Solomon B, Braun AR (2013): Adaptive significance of right hemisphere activation in aphasic language comprehension. Neuropsychologia 51:1248–1259.

Muhle-Karbe PS, Derrfuss J, Lynn MT, Neubert FX, Fox PT, Brass M, Eickhoff SB (2015): Co-Activation-Based Parcellation of the Lateral Prefrontal Cortex Delineates the Inferior Frontal Junction Area. Cereb Cortex:1–17.

van Oers CMM, Vink M, van Zandvoort MJE, van der Worp HB, de Haan EHF, Kappelle LJ, Ramsey NF, Dijkhuizen RM (2010): Contribution of the left and right inferior frontal gyrus in recovery from aphasia. A functional MRI study in stroke patients with preserved hemodynamic responsiveness. Neuroimage 49:885–93.

Oldfield RC (1971): the Assessment and Analysis of Handedness: the Edinburgh Inventory. Neuropsychologia 9:97–113.

Ovadia-Caro S, Villringer K, Fiebach J, Jungehulsing GJ, van der Meer E, Margulies DS, Villringer A (2013): Longitudinal effects of lesions on functional networks after stroke. J Cereb Blood Flow Metab 33:1279–85.

Pani E, Zheng X, Wang J, Norton A, Schlaug G (2016): Right hemisphere structures predict poststroke speech fluency. Neurology 86:1574–1581.

de Pasquale F, Della Penna S, Snyder AZ, Marzetti L, Pizzella V, Romani GL, Corbetta M (2012): A cortical core for dynamic integration of functional networks in the resting human brain. Neuron 74:753–64.

Pedersen P (1995): Aphasia in acute stroke: incidence, determinants, and recovery. Ann Neurol 38:659–666.

Piervincenzi C, Petrilli A, Marini A, Caulo M, Committeri G, Sestieri C (2016): Multimodal assessment of hemispheric lateralization for language and its relevance for behavior. Neuroimage.

Poldrack RA (2012): The future of fMRI in cognitive neuroscience. Neuroimage 62:1216–20.

Raboyeau G, De Boissezon X, Marie N, Balduyck S, Puel M, Bézy C, Démonet JF, Cardebat D (2008): Right hemisphere activation in recovery from aphasia: Lesion effect or function recruitment? Neurology 70:290-298.

Raichle ME, MacLeod AM, Snyder AZ, Powers WJ, Gusnard D a, Shulman GL (2001): A default mode of brain function. Proc Natl Acad Sci U S A 98:676–82.

Raichle ME, Snyder AZ (2007): A default mode of brain function: a brief history of an evolving idea. Neuroimage 37:1083–90; discussion 1097–9.

Ripollés P, Marco-Pallarés J, de Diego-Balaguer R, Miró J, Falip M, Juncadella M, Rubio F, Rodriguez-Fornells A (2012): Analysis of automated methods for spatial normalization of lesioned brains. Neuroimage 60:1296–306.

Rorden C, Karnath H-O (2004): Using human brain lesions to infer function: a relic from a past era in the fMRI age? Nat Rev Neurosci 5:813–819.

Rosen HJ, Petersen SE, Linenweber MR, Snyder AZ, White DA, Chapman L, Dromerick AW, Fiez JA, Corbetta MD (2000): Neural correlates of recovery from aphasia after damage to left inferior frontal cortex. Neurology 55:1883–1894.

Saur D, Hartwigsen G (2012): Neurobiology of language recovery after stroke: lessons from neuroimaging studies. Arch Phys Med Rehabil 93:S15-25.

Saur D, Lange R, Baumgaertner A, Schraknepper V, Willmes K, Rijntjes M, Weiller C (2006): Dynamics of language reorganization after stroke. Brain 129:1371–84.

Schwartz MF, Kimberg DY, Walker GM, Faseyitan O, Brecher A, Dell GS, Coslett HB (2009): Anterior temporal involvement in semantic word retrieval: voxel-based lesion-symptom mapping evidence from aphasia. Brain 132:3411–27.

Seghier ML, Ramlackhansingh A, Crinion J, Leff AP, Price CJ (2008): Lesion identification using unified segmentation-normalisation models and fuzzy clustering. Neuroimage 41:1253–1266.

Sestieri C, Corbetta M, Romani GL, Shulman GL (2011): Episodic memory retrieval, parietal cortex, and the default mode network: functional and topographic analyses. J Neurosci 31:4407–20.

Siegel JS, Ramsey LE, Snyder AZ, Metcalf N V, Chacko RV, Weinberger K, Baldassarre A, Hacker C, Shulman GL (2016): Common and specific disruptions of network connectivity predict impairment in multiple behavioral domains after stroke. PNAS I:1–10.

Sims JA, Kapse K, Glynn P, Sandberg C, Tripodis Y, Kiran S (2016): The relationships between the amount of spared tissue, percent signal change, and accuracy in semantic processing in aphasia. Neuropsychologia 84:113–126.

Slotnick SD, Moo LR, Segal JB, Hart J (2003): Distinct prefrontal cortex activity associated with item memory and source memory for visual shapes. Cogn Brain Res 17:75–82.

Szaflarski JP, Binder JR, Possing ET, McKiernan KA, Ward BD, Hammeke TA (2002): Language lateralization in left-handed and ambidextrous people: fMRI data. Neurology 59:238–244.

Szaflarski JP, Allendorfer JB, Banks C, Vannest J, Holland SK (2013): Recovered vs. not-recovered from post-stroke aphasia: the contributions from the dominant and non-dominant hemispheres. Restor Neurol Neurosci 31:347–60.

Szaflarski JP, Holland SK, Jacola LM, Lindsell C, Privitera MD, Szaflarski M (2008): Comprehensive presurgical functional MRI language evaluation in adult patients with epilepsy. Epilepsy Behav 12:74–83.

Szaflarski JP, Vannest J, Wu SW, DiFrancesco MW, Banks C, Gilbert DL (2011): Excitatory repetitive transcranial magnetic stimulation induces improvements in chronic post-stroke aphasia. Med Sci Monit 17:CR132-9.

Thiel A, Schumacher B, Wienhard K, Gairing S, Kracht LW, Wagner R, Haupt WF, Heiss W-D (2006): Direct demonstration of transcallosal disinhibition in language networks. J Cereb Blood Flow Metab 26:1122–7.

Turkeltaub PE, Coslett HB, Thomas AL, Faseyitan O, Benson J, Norise C, Hamilton RH (2012): The right hemisphere is not unitary in its role in aphasia recovery. Cortex 48:1179–1186.

Turkeltaub PE, Messing S, Norise C, Hamilton RH (2011): Are networks for residual language function and recovery consistent across aphasic patients? Neurology 76:1726–34.

Uddin LQ, Supekar K, Amin H, Rykhlevskaia E, Nguyen DA, Greicius MD, Menon V (2010): Dissociable connectivity within human angular gyrus and intraparietal sulcus: evidence from functional and structural connectivity. Cereb Cortex 20:2636–46.

Veldsman M, Cumming T, Brodtmann A (2015): Beyond BOLD: Optimizing functional imaging in stroke populations. Hum Brain Mapp 36:1620–1636.

Vincent JL, Kahn I, Snyder AZ, Raichle ME, Buckner RL (2008): Evidence for a frontoparietal control system revealed by intrinsic functional connectivity. J Neurophysiol 100:3328–42.

Wager TD, Lindquist M, Kaplan L (2007): Meta-analysis of functional neuroimaging data: Current and future directions. Soc Cogn Affect Neurosci 2:150–158.

Warren JE, Crinion JT, Lambon Ralph MA, Wise RJS (2009): Anterior temporal lobe connectivity correlates with functional outcome after aphasic stroke. Brain

132:3428–42.

Wilke M, de Haan B, Juenger H, Karnath H-O (2011): Manual, semi-automated, and automated delineation of chronic brain lesions: a comparison of methods. Neuroimage 56:2038–46.

Winhuisen L, Thiel A, Schumacher B, Kessler J, Rudolf J, Haupt WF, Heiss WD (2005): Role of the contralateral inferior frontal gyrus in recovery of language function in poststroke aphasia: a combined repetitive transcranial magnetic stimulation and positron emission tomography study. Stroke 36:1759–63.

Winhuisen L, Thiel A, Schumacher B, Kessler J, Rudolf J, Haupt WF, Heiss WD (2007): The right inferior frontal gyrus and poststroke aphasia: a follow-up investigation. Stroke 38:1286–92.

Xing S, Lacey EH, Skipper-Kallal LM, Jiang X, Harris-Love ML, Zeng J, Turkeltaub PE (2015): Right hemisphere grey matter structure and language outcomes in chronic left hemisphere stroke. Brain.

Xu Y, Lin Q, Han Z, He Y, Bi Y (2016): Intrinsic functional network architecture of human semantic processing: Modules and hubs. Neuroimage 132:542–555.

Yarkoni T (2009): Big Correlations in Little Studies: Inflated fMRI Correlations Reflect Low Statistical Power-Commentary on Vul et al. (2009). Perspect Psychol Sci 4:294–298.

Yarnell P, Monroe P, Sobel L (1976): Aphasia outcome in stroke: a clinical neuroradiological correlation. Stroke 7:516–522.

Zhang Y, Kimberg D, Coslett H, Schwartz M, Wang Z (2014): Multivariate lesion-symptom mapping using support vector regression. Hum Brain Mapp 35:997.

# Tables

**Table 1. Participant demographics**

| Group | N | Age | Sex | Handedness | Lesion volume (ml) |
|---|---|---|---|---|---|
| Patients | 43 | 53 (15) | 25 M | 0.85 (0.43) | 105.24 (76.29) |
| Controls | 43 | 54 (14) | 23 M | 0.80 (0.41) | N/A |

*Mean (SD) are shown for age/handedness; M-Male*

**Table 2. Cluster and peak statistics for control and patient SD activations**

| Group | Peak Location | Extent | t-value | x | y | z |
|---|---|---|---|---|---|---|
| Controls | L Superior Medial Gyrus | 44277 | 14.3308 | -4 | 46 | 36 |
| | R Posterior Cingulate | 44277 | 10.8963 | 2 | -34 | 32 |
| | L IFG (p. Orbitalis) | 44277 | 10.6975 | -36 | 30 | -8 |
| | L Angular Gyrus | 3981 | 11.8496 | -40 | -64 | 36 |
| | R Angular Gyrus | 1051 | 8.1312 | 50 | -60 | 32 |
| | R Mid Orbital Gyrus | 125 | 3.3213 | 2 | 66 | -12 |
| Patients | R Cerebelum (VIII) | 3519 | 5.1981 | 10 | -76 | -28 |
| | R Cerebelum (IX) | 3519 | 4.1282 | 8 | -54 | -52 |
| | R Cerebelum (VII) | 3519 | 3.8702 | 42 | -66 | -48 |
| | L Superior Medial Gyrus | 7349 | 4.9037 | -2 | 40 | 56 |
| | L IFG (p. Triangularis) | 7349 | 4.7172 | -48 | 28 | 8 |
| | R Superior Frontal Gyrus | 7349 | 4.0744 | 16 | 54 | 30 |
| | L Middle Temporal Gyrus | 253 | 4.409 | -52 | -8 | -18 |
| | L Middle Temporal Gyrus | 1085 | 4.4082 | -44 | -64 | 28 |
| | R Angular Gyrus | 254 | 4.0334 | 54 | -68 | 32 |
| | Posterior Cingulate | 217 | 3.9907 | 0 | -32 | 30 |
| | L Inferior Temporal Gyrus | 585 | 3.9528 | -60 | -32 | -12 |
| | R Fusiform Gyrus | 111 | 3.7833 | 40 | -24 | -28 |
| | R Posterior Cingulate | 265 | 2.9714 | 4 | -46 | 14 |

*Note: Cluster peaks are provided for the top 3 peaks/cluster that are separated by a minimum distance of 30mm.*

**Table 3. Cluster and peak statistics for correlations between patient SD activation and language task performance while controlling for lesion volume.**

| Sign | Peak Location | Extent | t-value | x | y | z |
|---|---|---|---|---|---|---|
| | **Semantic Decisions** | | | | | |
| Positive | L Angular Gyrus | 1913 | 4.9167 | -42 | -68 | 48 |
| | L Middle Temporal Gyrus | 1913 | 2.5892 | -50 | -46 | 8 |
| | Cerebellar Vermis (VII) | 5291 | 4.8921 | -2 | -78 | -18 |
| | Posterior Cingulate Gyrus | 5291 | 4.4726 | 0 | -32 | 32 |
| | L Calcarine Gyrus | 5291 | 4.3329 | 4 | -94 | 12 |
| | Dorsal Anterior Pons | 544 | 4.6492 | 0 | -12 | -28 |
| | L Middle Orbital Gyrus | 961 | 4.5543 | -42 | 54 | 0 |
| | L Middle Temporal Gyrus | 785 | 4.4425 | -68 | -32 | -2 |
| | L Inferior Temporal Gyrus | 671 | 4.3857 | -32 | -6 | -38 |
| | R Angular Gyrus | 547 | 4.2227 | 54 | -58 | 40 |
| | L Parahippocampal Gyrus | 457 | 4.0258 | -22 | -28 | -18 |
| | R Parahippocampal Gyrus | 275 | 3.9305 | 20 | -32 | -20 |
| | L Superior Frontal Gyrus | 175 | 3.745 | -20 | 56 | 32 |
| | R IFG (p. Orbitalis) | 160 | 3.7179 | 50 | 38 | -18 |
| | R Precuneus | 211 | 3.6912 | 4 | -62 | 66 |
|  | R Rectal Gyrus | 135 | 3.6875 | 2 | 24 | -30 |
|  | R Inferior Temporal Gyrus | 153 | 3.6576 | 56 | -12 | -30 |
|  | R Temporal Pole | 421 | 3.6249 | 36 | 14 | -24 |
| Negative | L Postcentral Gyrus | 258 | -3.4808 | -60 | -10 | 30 |
| **Fluency** | | | | | | |
| Positive | R Superior Medial Gyrus | 7588 | 6.0189 | 2 | 50 | 38 |
|  | L Middle Frontal Gyrus | 7588 | 5.8247 | -34 | 18 | 52 |
|  | L Posterior-Medial Frontal | 7588 | 5.1085 | -4 | 20 | 66 |
|  | L Brainstem (Inf. Colliculus?) | 6302 | 5.2447 | -2 | -30 | -20 |
|  | L Fusiform Gyrus | 6302 | 4.9057 | -34 | -18 | -30 |
|  | R Cerebelum (IX) | 6302 | 4.8778 | 8 | -56 | -38 |
|  | L Angular Gyrus | 1227 | 5.2081 | -48 | -62 | 32 |
|  | L Posterior Cingulate | 4805 | 4.6649 | -2 | -46 | 18 |
|  | R Cerebelum (VI) | 4805 | 4.0471 | 28 | -80 | -14 |
|  | L Calcarine Gyrus | 4805 | 3.6055 | 0 | -92 | 6 |
|  | R Fusiform Gyrus | 232 | 4.616 | 34 | -18 | -34 |
|  | R Medial Temporal Pole | 1532 | 4.5911 | 46 | 6 | -36 |
|  | R Middle Temporal Gyrus | 1532 | 3.7147 | 68 | -16 | -20 |
|  | R Parahippocampal Gyrus | 1532 | 3.685 | 16 | -2 | -30 |
|  | L IFG (p. Orbitalis) | 587 | 4.0197 | -40 | 24 | -12 |
| Negative | R Middle Temporal Gyrus | 155 | -3.3541 | 48 | -48 | 2 |
| **Naming** | | | | | | |
| Positive | R Superior Medial Gyrus | 5204 | 5.5202 | 2 | 54 | 46 |
|  | L Middle Frontal Gyrus | 5204 | 4.494 | -34 | 16 | 52 |
|  | R Posterior-Medial Frontal | 5204 | 3.4022 | 6 | 14 | 74 |
|  | L Angular Gyrus | 1889 | 4.2896 | -46 | -60 | 30 |
|  | R Inferior Temporal Gyrus | 268 | 4.207 | 48 | 2 | -34 |
|  | L Fusiform Gyrus | 474 | 3.9476 | -40 | -34 | -14 |
|  | R IFG (p. Orbitalis) | 354 | 3.921 | 44 | 24 | -14 |
|  | R Fusiform Gyrus | 265 | 3.9092 | 16 | -8 | -42 |
|  | R Superior Orbital Gyrus | 265 | 2.6356 | 16 | 14 | -18 |
|  | R Cerebelum (IX) | 762 | 3.8878 | 14 | -46 | -38 |
|  | R Precuneus | 2516 | 3.8798 | 2 | -66 | 64 |
|  | Posterior Cingulate | 2516 | 3.5849 | 0 | -36 | 32 |
|  | L Superior Parietal Lobule | 2516 | 2.4993 | -28 | -56 | 66 |
|  | L IFG (p. Orbitalis) | 202 | 3.8209 | -42 | 26 | -16 |
|  | L Fusiform Gyrus | 139 | 3.6031 | -18 | 6 | -38 |

*Note: Cluster peaks are provided for the top 3 peaks/cluster that are separated by a minimum distance of 30mm*

**Table 4. Cluster and peak statistics for right hemispheric correlations between SD activation and lesion volume.**

| Sign | Peak Location | Extent | t-value | x | y | z |
|---|---|---|---|---|---|---|
| Positive | R IFG (p. Opercularis) | 277 | 3.9446 | 56 | 12 | 8 |
|  | R Posterior-Medial Frontal | 200 | 3.0469 | 8 | 6 | 56 |
|  | R Middle Frontal Gyrus | 143 | 3.0383 | 34 | 40 | 20 |
|  | R Precentral Gyrus | 221 | 2.9707 | 54 | -2 | 40 |
| Negative | R Inferior Temporal Gyrus | 399 | -4.1944 | 56 | -14 | -34 |
|  | R Mid Orbital Gyrus | 223 | -3.9815 | 2 | 46 | 4 |
|  | R Cerebelum (Crus I) | 1137 | -3.8737 | 56 | -62 | -32 |
|  | R Cerebelum (IX) | 1137 | -3.3749 | 14 | -56 | -48 |
|  | R Cerebelum (Crus I) | 1137 | -3.1767 | 34 | -82 | -24 |
|  | R Anterior Thalamus/Putamen | 160 | -3.7215 | 10 | -2 | -6 |

*Note: Cluster peaks are provided for the top 3 peaks/cluster that are separated by a minimum distance of 30mm*

**Figures**

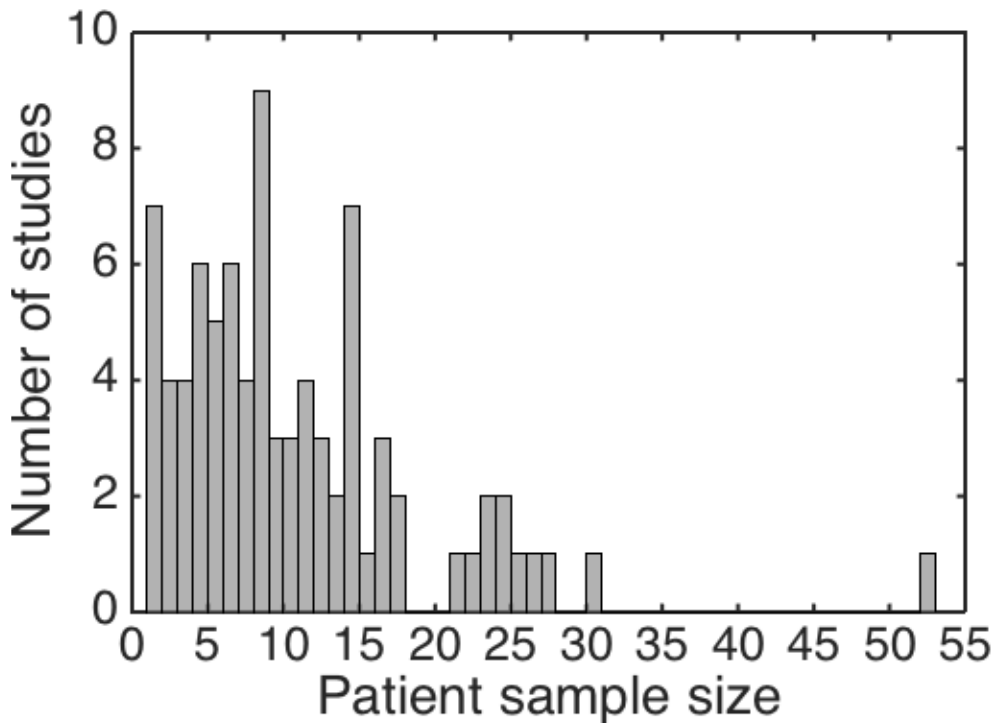

**Figure 1. Sample sizes in the functional neuroimaging literature of post-stroke aphasia**. Histogram illustrating frequencies of different sample sizes across 84 functional neuroimaging studies of aphasia published since 1995. Bins correspond to single integer values. The average sample size was 10.33, with a standard deviation of 8.44. The single largest sample was 53 patients. 67 (81%) of studies had samples sizes less than or equal to 15 patients, and 73 (87%) of studies had sample sizes less than or equal to 20 patients.

Individual study sample sizes and neuroimaging modalities are provided in Supplementary Table 1, and the full reference list is also provided in Supplementary Material 1.

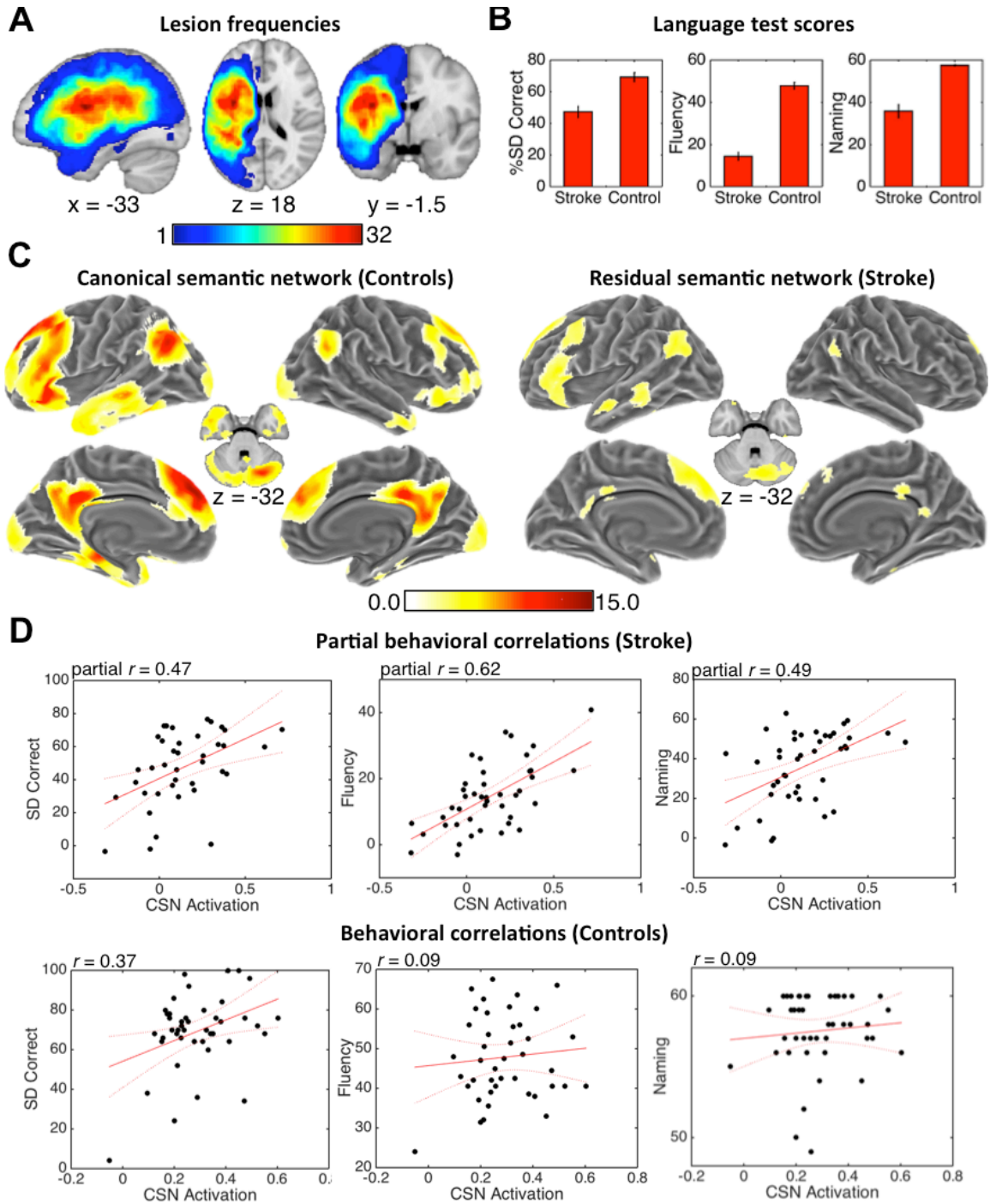

**Figure 2. Data characterization and partial behavioral correlation results. A.** Lesion frequencies across all 43 patients. Minimum colorbar values indicate voxels lesioned in only 1 patient, and maximum colorbar values indicate voxels lesioned in 32 patients (maximum lesion overlap). **B.** Means and standard errors for stroke patients and healthy controls are shown for each language measure. **C.** Areas showing significantly more activity during the SD condition relative to the control condition in the healthy controls (left) and areas showing significantly more activity during the SD condition relative to the TD condition in the stroke patients (right). Both maps are intensity thresholded at $p<0.01$, uncorrected and cluster-corrected at $p<0.05$ (99 voxels). Colorbar values indicate t-statistics. Note that the canonical semantic network (CSN) region of interest (ROI) was defined based on the results shown in (C, left). **D.** Scatterplots illustrate the partial correlations between the residuals for activation in the CSN ROI and the residuals for performance on each language measure (y-axes) after removing the effects of lesion volume for the patient group (top). Scatterplots illustrate correlations between activation in the CSN ROI and performance on each language measure for the control group (bottom). The line of best fit (red) and 95% confidence intervals (red dashes) are shown on each plot. Each relationship shown was significant at $p<0.05$, family-wise error corrected. Note: All brain renderings are in neurological convention.

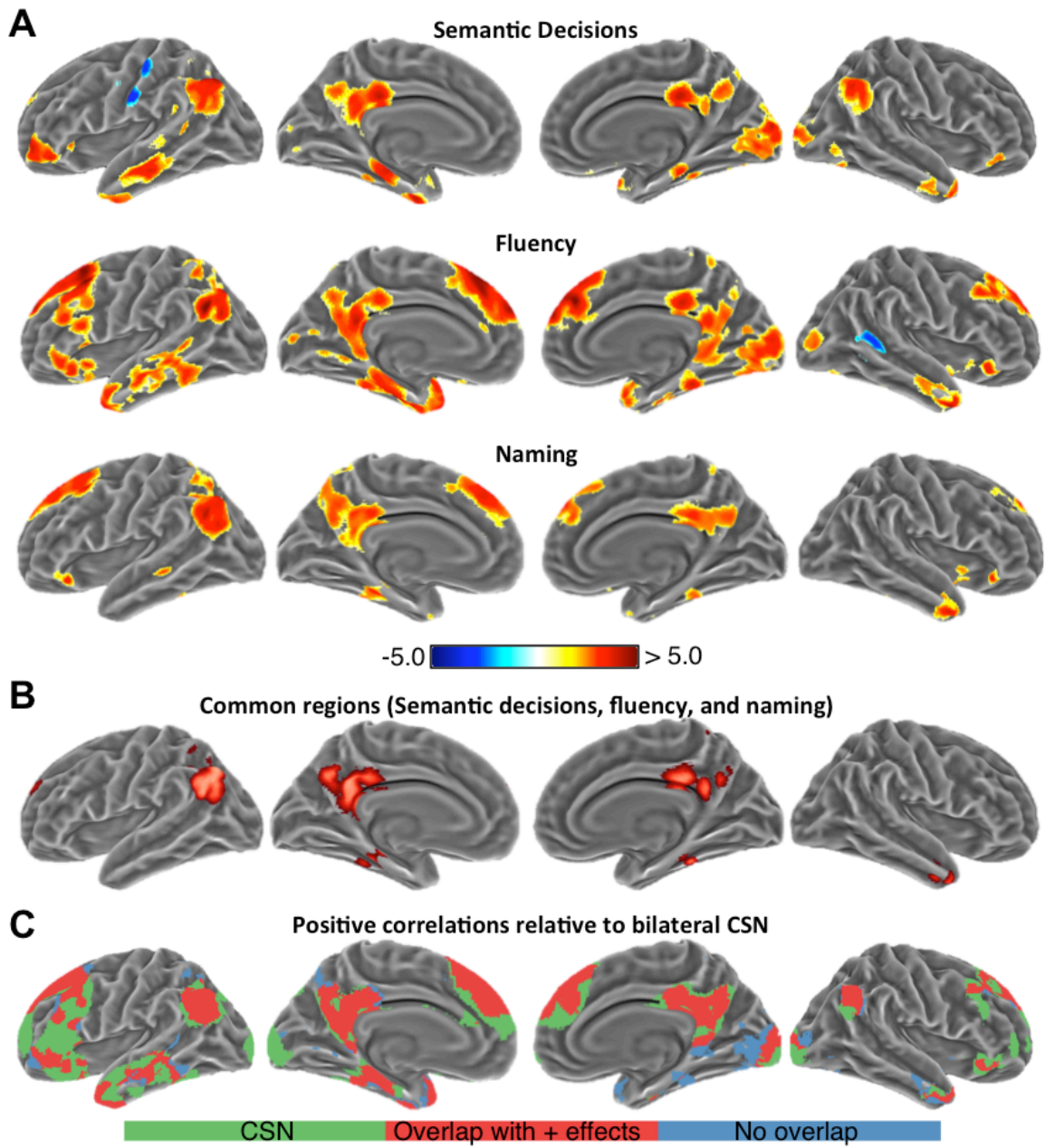

**Figure 3. Whole-brain performance regression results. A.** Positive correlations (hot colors) and negative correlations (cold colors) identified for the patient group between task-driven activation and performance on the SD task (top) combined fluency measure (middle) and Boston Naming Test (bottom). **B.** The subset of regions showing positive correlations between SD activation and performance on all language measures in the patient group are shown in red. **C**. The canonical network identified in controls is shown in green, regions within the CSN identified in healthy controls where SD activation was positively related to performance on any language measure in patients are shown in red,

and regions outside of the CSN where SD activation was positively related to performance on any language measure in patients are shown in blue. Note that the overlays shown in (C) are qualitative illustrations of how the quantitatively identified behavioral relationships in patients relate to the canonical network identified in controls, and are intended to illustrate how activity supporting residual language task performance relates to the CSN. Each map is intensity thresholded at $p<0.01$, uncorrected and cluster-corrected at $p<0.01$ (126 voxels). Colorbar values indicate t-statistics. All brain renderings are in neurological convention.

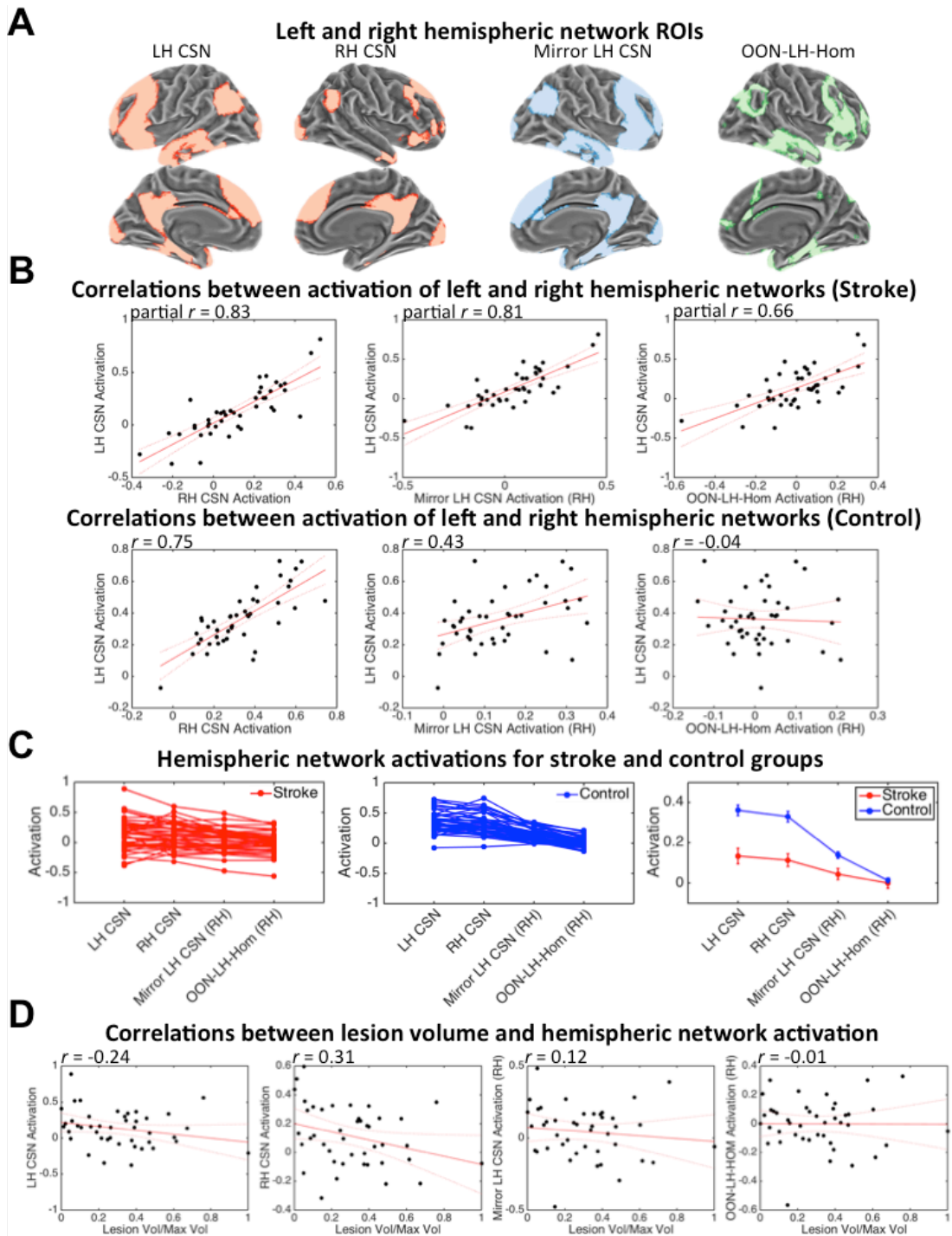

**Figure 4. Relationships between left and right hemispheric activation. A**. Brain renderings show the ROIs corresponding to the left hemispheric CSN (LH CSN), right hemispheric CSN (RH CSN), mirrored left hemispheric CSN (Mirror LH CSN), and out-of-network left hemispheric homologues (OON-LH-Hom). **B.** Scatterplots illustrating the relationship between residual (with the effects of lesion volume partialled out) SD-driven

activity in the LH CSN ROI (y-axes) and each right hemispheric ROI (x-axes) are shown for the stroke group (top). Scatterplots illustrating the relationship between SD-driven activity in the LH CSN ROI (y-axes) and each right hemispheric ROI (x-axes) are also shown for the control group (bottom). **C.** Line plots illustrate relative activation magnitudes in the LH CSN ROI and each right hemispheric ROI for the stroke (left, red) and control (middle, blue) groups. Means and standard errors of activation magnitudes at each ROI are also shown for both groups (right). **D.** Scatterplots illustrate the relationships between left hemispheric lesion volume and activity in each ROI.

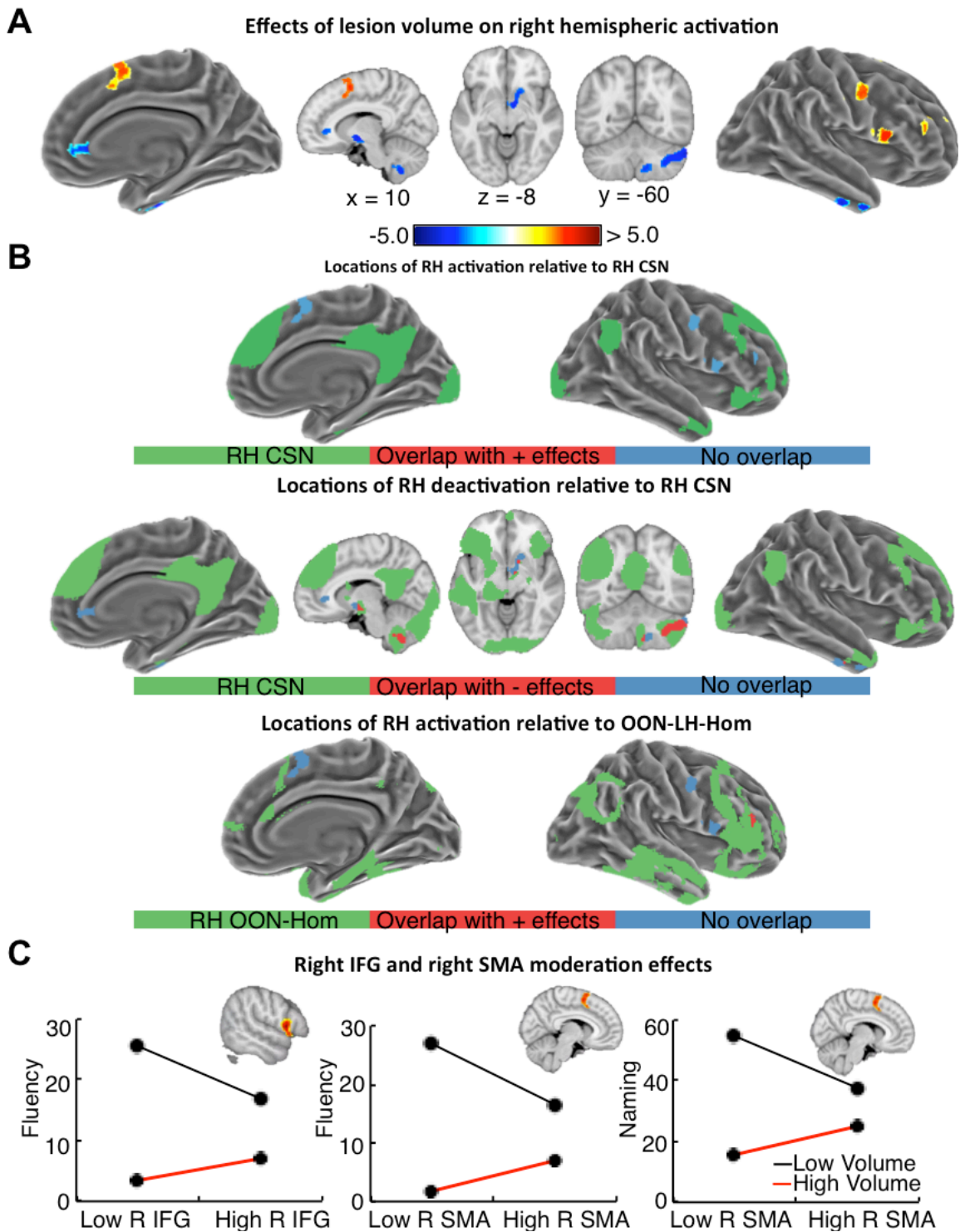

**Figure 5. Lesion effects on regional activation in the right hemisphere. A.** Positive correlations (hot colors) and negative correlations (cold colors) between task-driven activation and lesion volume in right hemispheric areas are shown (top). **B.** Areas of overlap (red) and no overlap (blue) between regions where increased activation is associated with larger left hemispheric lesion volumes and the right hemispheric CSN ROI are shown on the top row. Areas of overlap (red) and no overlap (blue) between regions where decreased activation is associated with larger left hemispheric lesion volumes and the right hemispheric CSN ROI are shown on the middle row. Areas of overlap (red) and no overlap (blue) between regions where increased activation is associated with larger left hemispheric lesion volumes and the out-of-network left hemispheric homologues (OON-LH-Hom -- green) ROI are shown on the bottom row. Note that the overlays shown in (B) are qualitative illustrations of the quantitatively identified lesion volume effects on right hemispheric activations, and illustrate how right hemispheric activations associated with lesion volume effects relate to the RH CSN ROI (top, middle) and to the OON-LH-Hom ROI (bottom). **C**. Interaction plots are shown for the moderating effects of lesion volume on the relationship between right inferior frontal gyrus (IFG) activation and fluency scores (left), right supplementary motor area (SMA) activation and fluency scores (middle), and right SMA activation and naming scores (right). Each of these relationships survived a False Discovery Rate threshold of 0.1. High and low values shown in the interaction plots correspond to +1 and -1 standard deviations from the mean of each variable. All maps are intensity thresholded at $p<0.01$, uncorrected and cluster-corrected at $p<0.01$ (126 voxels). Colorbar values indicate t-statistics. Note: all brain renderings are in neurological convention.

**Supplementary Material**

**Supplementary Analysis 1. Sample sizes from literature search**

**Supplementary Table 1. Aphasia patient sample sizes in 84 functional neuroimaging studies of aphasia between 1995 and 2016.**

| Study | N | Modality | Study | N | Modality |
|---|---|---|---|---|---|
| **Weiller et al., 1995** | 6 | PET | Abutalebi et al., 2009 | 1 | fMRI |
| **Belin et al., 1996** | 7 | PET | Breier et al., 2009 | 23 | MEG |
| **Cappa et al., 1997** | 8 | PET | Fridriksson et al., 2009 | 11 | fMRI |
| **Heiss et al., 1997** | 6 | PET | Martin et al., 2009 | 2 | fMRI |
| **Thomas et al., 1997** | 11 | EEG | Specht et al., 2009 | 12 | PET |
| **Sakatani et al., 1998** | 10 | NIRS | Warren et al., 2009 | 24 | PET |
| **Karbe et al., 1998** | 7 | PET | Fridriksson et al., 2010a | 15 | fMRI |
| **Heiss et al., 1999** | 23 | PET | Fridriksson et al., 2010b | 26 | fMRI |
| **Miura et al., 1999** | 1 | fMRI | Postman-Caucheteux et al., 2010 | 3 | fMRI |
| **Musso et al., 1999** | 4 | PET | Rochon et al., 2010 | 4 | fMRI |
| **Warburton et al., 1999** | 6 | PET | Saur et al., 2010 | 21 | fMRI |
| **Rosen et al., 2000** | 6 | PET+fMRI | Sharp et al., 2010 | 9 | PET |
| **Blasi et al., 2002** | 8 | fMRI | Thompson et al., 2010a | 5 | fMRI |
| **Leger et al., 2002** | 1 | fMRI | Thompson et al., 2010b | 6 | fMRI |
| **Blank et al., 2003** | 14 | PET | Van Oers et al., 2010 | 13 | fMRI |
| **Cardebat et al., 2003** | 8 | PET | Papoutsi et al., 2011 | 14 | fMRI |
| **Perani et al., 2003** | 5 | fMRI | Szaflarski et al., 2011a | 4 | fMRI |
| **Fernandez et al., 2004** | 1 | fMRI | Szaflarski et al., 2011b | 8 | fMRI |
| **Kurland et al., 2004** | 2 | fMRI | Tyler et al., 2011 | 14 | fMRI |
| **Naeser et al., 2004** | 4 | fMRI | Weiduschat et al., 2011 | 10 | PET |
| **Peck et al., 2004** | 3 | fMRI | Allendorfer et al., 2012 | 16 | fMRI |
| **Zahn et al., 2004** | 7 | fMRI | Fridriksson et al., 2012 | 30 | fMRI |
| **Crinion et al., 2005** | 17 | fMRI | Meltzer et al., 2013 | 25 | MEG |
| **Crosson et al., 2005** | 2 | fMRI | Szaflarski et al., 2013 | 27 | fMRI |
| **De Boissezon et al., 2005** | 7 | PET | Thompson et al., 2013 | 8 | fMRI |
| **Martin et al., 2005** | 5 | fMRI | Thiel et al., 2013 | 24 | PET |
| **Pulvermuller et al., 2005** | 9 | EEG | Abel et al., 2014 | 14 | fMRI |
| **Winhuisen et al., 2005** | 11 | PET | Brownsett et al., 2014 | 16 | fMRI |
| **Meinzer et al., 2006** | 1 | fMRI | Jarso et al., 2014 | 4 | fMRI |
| **Connor et al., 2006** | 8 | fMRI | Mattioli et al., 2014 | 12 | fMRI |
| **Crinion et al., 2006** | 22 | PET | Mohr et al., 2014 | 6 | fMRI |
| **Fridriksson et al., 2006** | 3 | fMRI | Robson et al., 2014 | 12 | fMRI |
| **Saur et al., 2006** | 14 | fMRI | Seghier et al., 2014 | 1 | fMRI |

| Bonakdarpour et al., 2007 | 5 | fMRI | Van Hees et al., 2014 | 8 | fMRI |
|---|---|---|---|---|---|
| Fridriksson et al., 2007 | 3 | fMRI | Zhu et al., 2014 | 13 | fMRI |
| Meinzer et al., 2007 | 1 | fMRI | Abel et al., 2015 | 14 | fMRI |
| Winhuisen et al., 2007 | 9 | PET | Bonakdarpour et al., 2015 | 5 | fMRI |
| Vitali et al., 2007 | 2 | fMRI | Kiran et al., 2015 | 8 | fMRI |
| Eaton et al., 2008 | 4 | fMRI | Spironelli et al., 2015 | 17 | EEG |
| Meinzer et al., 2008 | 11 | fMRI | Griffis et al., 2016 | 8 | fMRI |
| Raboyeau et al., 2008 | 10 | PET | Sims et al., 2016 | 14 | fMRI |
| Richter et al., 2008 | 16 | fMRI | Geranmayeh et al., 2016 | 53 | fMRI |

*References for supplementary Table 1 are at the end of the supplemental material.

**Supplementary Analysis 2. Additional characterization of patient data**

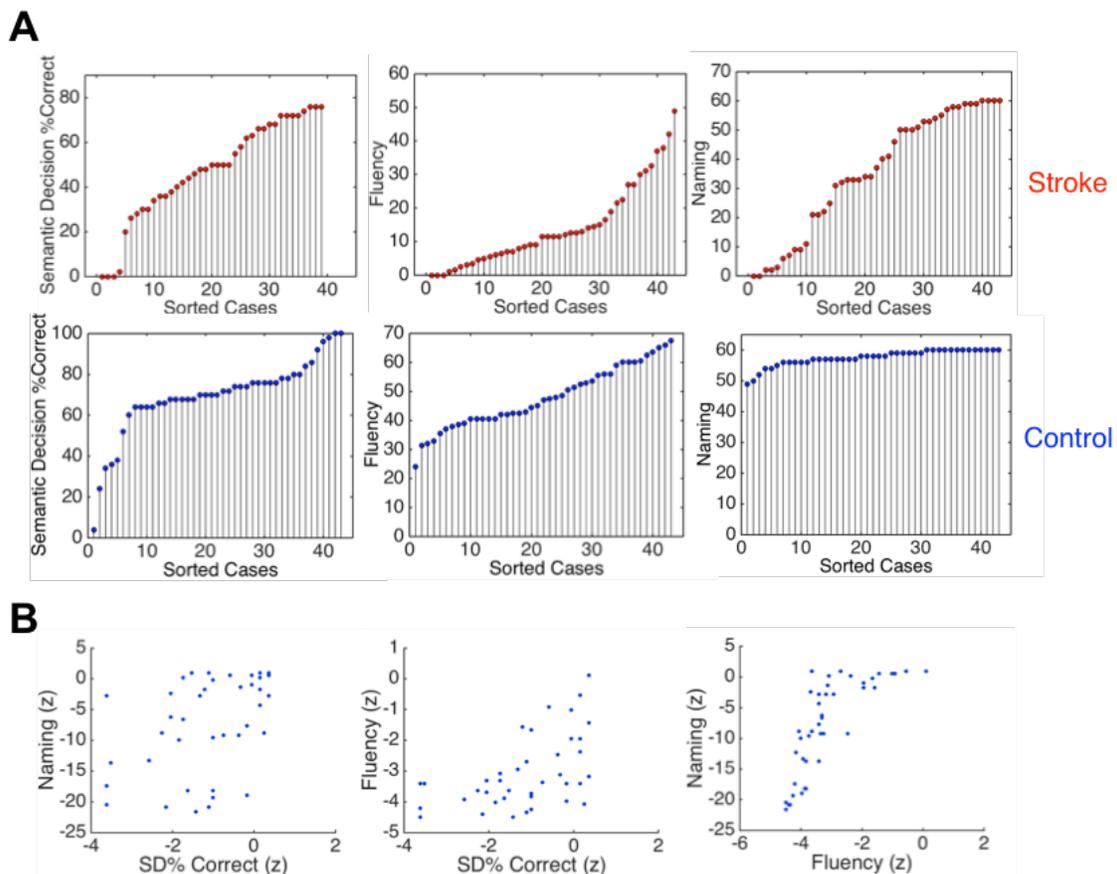

**Supplementary Figure 1. Behavioral measures. A.** Plots of ordered scores for semantic decisions (left), averaged fluency (middle), and picture naming (right). Scores are shown separately for stroke patients (top, red) and controls (bottom, blue). **B.** Plots of relationships between patient scores on (1) naming and semantic decisions (left), (2) fluency and semantic decisions (middle), and (3) fluency and naming (right). For each plot, patient scores are shown as z-scores based on the control data.

**Supplementary Table 2. Patient characteristics**

| Patient | Age | Sex | EHI | TSS | BNT | COWAT | SFT | Average Fluency | %SD Correct |
|---|---|---|---|---|---|---|---|---|---|
| 1 | 63 | F | 0.55 | 1.0 | 59 | 27 | 47 | 37 | 68 |
| 2 | 78 | F | 1.00 | 4.1 | 57 | 21 | 39 | 30 | 50 |
| 3 | 41 | F | 0.50 | 5.8 | 9 | 4 | 9 | 6.5 | 38 |
| 4 | 54 | M | 1.00 | 1.6 | 7 | 5 | 6 | 5.5 | 66 |
| 5 | 46 | M | 0.90 | 1.0 | 53 | 20 | 42 | 31 | 46 |
| 6 | 52 | M | 0.58 | 1.0 | 60 | 11 | 27 | 19 | 48 |
| 7 | 56 | M | 1.00 | 3.4 | 32 | 2 | 14 | 8 | 50 |
| 8 | 53 | M | 1.00 | 5.0 | 50 | 8 | 20 | 14 | 76 |
| 9 | 55 | M | 1.00 | 1.2 | 58 | 15 | 30 | 22.5 | 72 |
| 10 | 48 | M | 1.00 | 6.1 | 22 | 0 | 12 | 6 | 20 |
| 11 | 63 | M | 1.00 | 1.0 | 60 | 8 | 10 | 9 | 40 |
| 12 | 56 | F | 1.00 | 1.0 | 33 | 6 | 20 | 13 | NA |
| 13 | 23 | M | 1.00 | 1.0 | 60 | 36 | 62 | 49 | 76 |
| 14 | 50 | M | 1.00 | 1.0 | 2 | 0 | 2 | 1 | 28 |
| 15 | 48 | F | 1.00 | 1.0 | 60 | 24 | 60 | 42 | 72 |
| 16 | 70 | F | 1.00 | 2.0 | 11 | 3 | 3 | 3 | 0 |
| 17 | 68 | M | 0.91 | 3.3 | 9 | 4 | 10 | 7 | 50 |
| 18 | 59 | M | 0.82 | 1.0 | 53 | 19 | 35 | 27 | 72 |
| 19 | 23 | F | 1.00 | 1.0 | 59 | 20 | 45 | 32.5 | 76 |
| 20 | 24 | F | 1.00 | 1.0 | 59 | 31 | 45 | 38 | 58 |
| 21 | 78 | F | 1.00 | 3.4 | 58 | 9 | 21 | 15 | 36 |
| 22 | 65 | M | 1.00 | 14.0 | 55 | 18 | 36 | 27 | 68 |
| 23 | 58 | F | 1.00 | 13.0 | 40 | 13 | 12 | 12.5 | 36 |
| 24 | 72 | F | 1.00 | 1.5 | 0 | 0 | 0 | 0 | NA |
| 25 | 50 | M | 1.00 | 2.9 | 0 | 0 | 0 | 0 | 42 |
| 26 | 57 | M | 1.00 | 2.1 | 2 | 2 | 1 | 1.5 | 48 |
| 27 | 51 | M | 1.00 | 1.1 | 37 | 8 | 15 | 11.5 | 66 |
| 28 | 43 | M | 1.00 | 1.3 | 50 | 11 | 22 | 16.5 | 44 |
| 29 | 24 | M | 0.83 | 2.3 | 21 | 9 | 14 | 11.5 | 2 |
| 30 | 67 | F | 1.00 | 2.2 | 6 | 2 | 3 | 2.5 | 50 |
| 31 | 62 | F | -1.00 | 4.4 | 33 | 20 | 23 | 21.5 | 62 |
| 32 | 44 | F | 0.91 | 2.1 | 41 | 7 | 18 | 12.5 | 30 |
| 33 | 62 | M | 1.00 | 2.6 | 54 | 10 | 19 | 14.5 | 63 |
| 34 | 31 | M | 1.00 | 4.8 | 21 | 7 | 7 | 7 | NA |
| 35 | 61 | M | 1.00 | 9.6 | 25 | 1 | 6 | 3.5 | NA |
| 36 | 64 | M | -1.00 | 2.7 | 51 | 1 | 16 | 8.5 | 30 |
| 37 | 38 | F | 0.91 | 1.8 | 46 | 11 | 12 | 11.5 | 72 |
| 38 | 53 | F | 1.00 | 9.2 | 34 | 2 | 7 | 4.5 | 74 |
| 39 | 54 | M | 0.92 | 3.3 | 33 | 5 | 19 | 12 | 55 |

| | | | | | | | | |
|---|---|---|---|---|---|---|---|---|
| 40 | 46 | M | 1.00 | 1.3 | 31 | 1 | 9 | 5 | 34 |
| 41 | 90 | F | 0.71 | 1.3 | 3 | 0 | 0 | 0 | 0 |
| 42 | 29 | F | 1.00 | 3.4 | 50 | 4 | 19 | 11.5 | 0 |
| 43 | 67 | M | 1.00 | 12.4 | 34 | 2 | 16 | 9 | 26 |

*EHI – Edinburgh Handedness Inventory, TSS – time since stroke, BNT – Boston Naming Test, COWAT – Controlled Oral Word Association Test, SFT – Semantic Fluency Test, %SD Correct -- % Semantic Decision Correct.

*Note:* The average fluency score is the average of the COWAT and SFT.

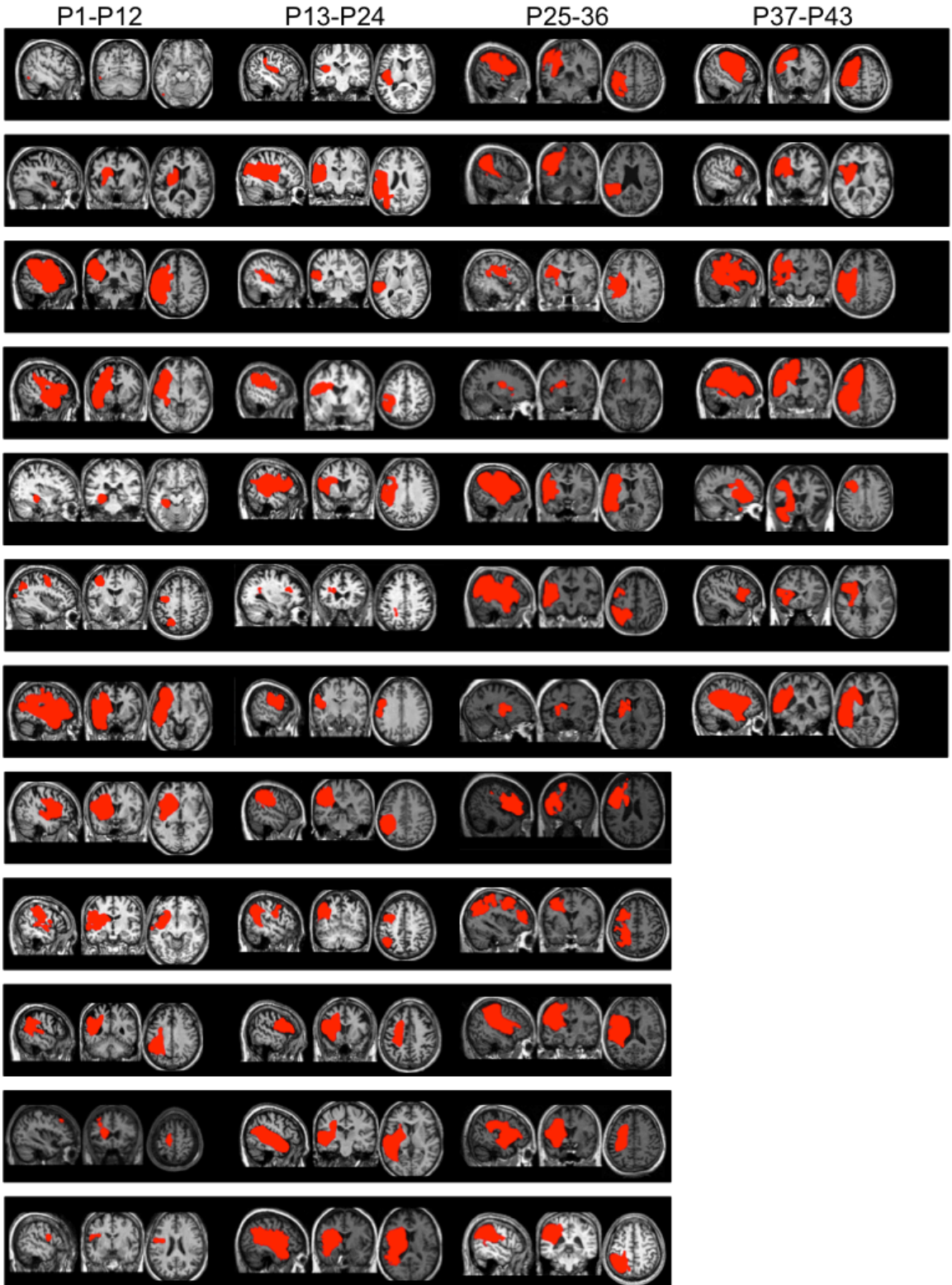

**Supplementary Figure 2.** Patient lesions. Patient lesions are shown on example slices. Patients are organized in columns, as indicated by the text above each column.

**Supplementary Analysis 3. Evaluation of result dependence on use of lesion volume and scanner covariates**

While our main analyses were performed using lesion volume as a covariate, additional analyses were performed to assess the stability of our results with and without scanner included as a covariate (since patients were collected on 2 different scanners) and with and without lesion volume as a covariate (where applicable). These analyses were performed to asses the effects of including scanner and lesion volume covariates on group results. Patients 1-20 were collected on the Phillips scanner, while patients 21-43 were collected on the Siemens scanner. We note that median lesion volumes significantly differed between patients collected on each scanner (Mann-Whitney U-test p=0.027), and mean lesion volumes showed a trend-level difference ($t_{41} = 1.83$, p=0.075) between scanners. Control participants were all collected on the Phillips scanner.

**Scanner and lesion volume covariates:** Mean canonical semantic network (CSN) activation still predicted fluency (partial r =0.56, p<0.001), naming (partial r=0.49, p=0.001), and SD performance (partial r=0.46, p=0.006) when both scanner and lesion volume were partialled out. Activity in left and right hemispheric portions of the CSN remained significantly correlated when both lesion volume and scanner were partialled out (partial r=0.79, p<0.001). The same was true when the mirror ROI was used (partial r=0.80, p<0.001), and when the out-of-network homologue ROI was used (r=0.70, p<0.001). Whole-brain regression results for models that included both scanner and lesion volume as covariates are shown in Supplementary Figure 3A.

**Scanner covariate only:** Mean CSN activation still predicted fluency (partial r = 0.56, p<0.001), naming (partial r=0.51, p<0.001), and SD performance (partial r=0.46, p=0.004) when only scanner was partialled out. Activity in the whole CSN remained negatively but non-significantly correlated with total lesion volume when scanner was partialled out (r=-0.17, p=0.28). Activity in left and right hemispheric portions of the CSN remained significantly correlated when only scanner was partialled out (partial r=0.79, p<0.001). The same was true when the mirror ROI was used (partial r=0.80, p<0.001), and when the out-of-network homologue ROI was used (r=0.69, p<0.001). Left CSN (partial r = -0.14, p=0.38) and right CSN (partial r=-0.22, p=0.17) activation remained negatively, but less significantly, correlated with lesion volume when scanner was partialled out. Whole-brain regression results for models that included only scanner as a covariate are shown in Supplementary Figure 3B.

**No covariates**: Correlational results are provided in the main text. Whole-brain regression results for models that did not include either scanner or lesion volume as covariates are shown in Figure 3C.

We note that results were highly similar for all analyses, indicating that scanner and lesion volume covariates had little effect on the results. While the minor effects of scanner are perhaps not surprising due to the use of identical scan parameters, our results suggest that lesion volume differences had only minor effects on functional MRI results in this sample.

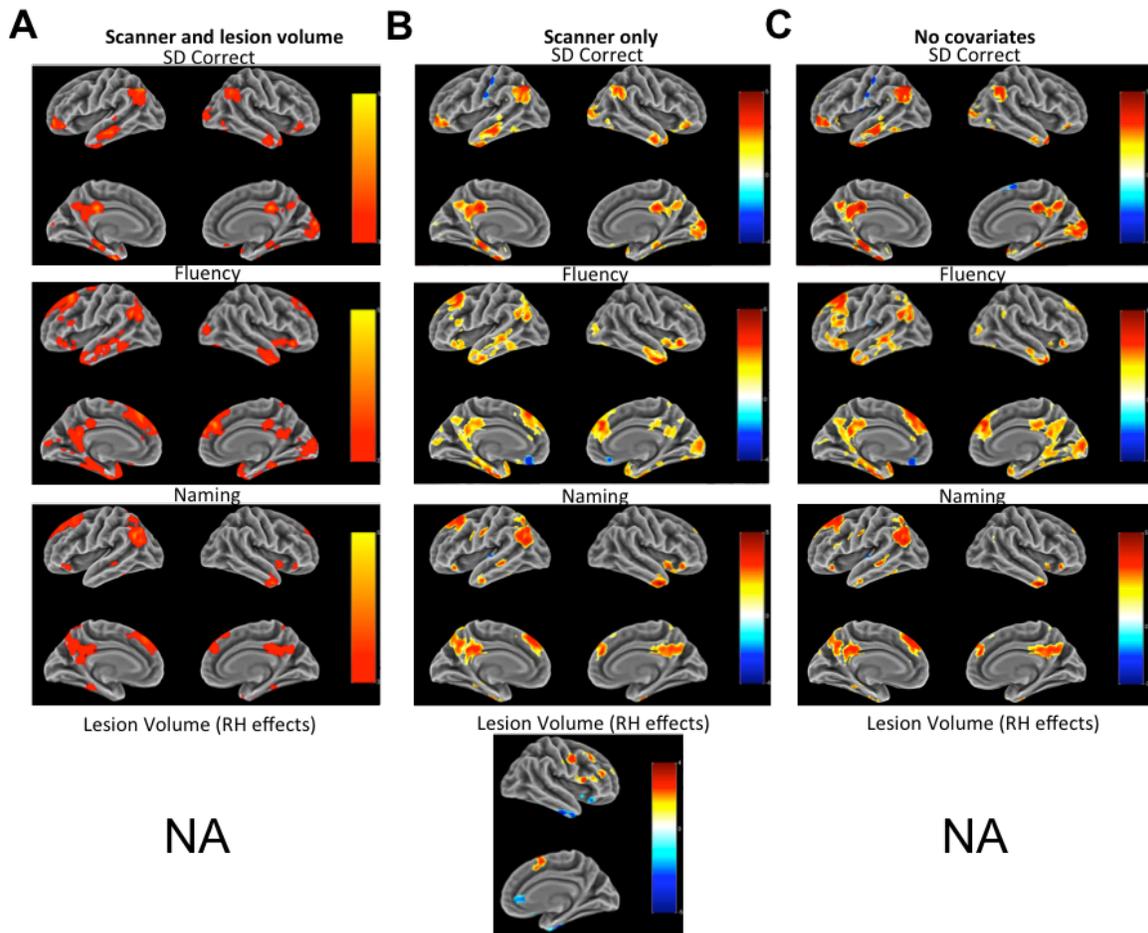

**Supplementary Figure 3**. Group analyses with and without different covariates included in the models. A. Whole-brain regressions with both scanner and lesion volume included as covariates. **B.** Whole-brain and right-hemispheric regressions with scanner included as a covariate. **C.** Whole-brain regressions with no covariates. Note: RH lesion volume effects are only shown for (B) because lesion volume is the variable of interest in this analysis, and thus scanner is the only applicable covariate.

Statistical maps are thesholded at a voxelwise p<0.01 and cluster-corrected at p<0.01 (126 voxels). Importantly, nearly identical results were obtained with and without each covariate included in the models.

**Supplementary Analysis 4. Semantic decision activation peaks for voxel-wise (FDR) correction thresholds.**

A

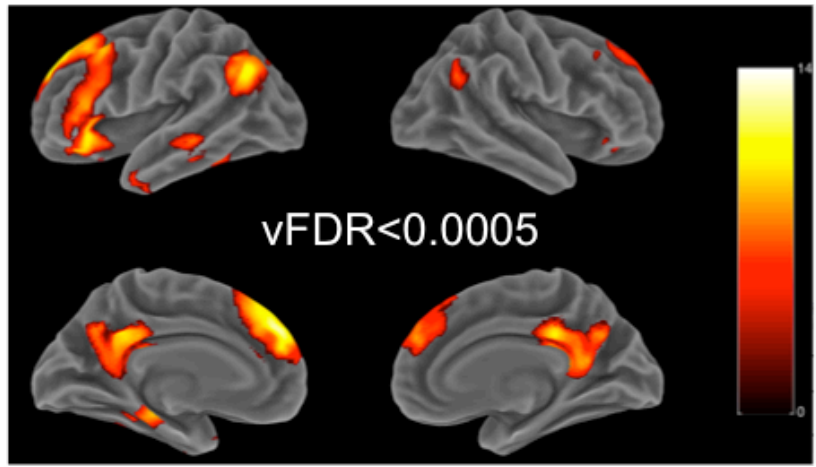

vFDR<0.0005

| Region | Cluster Size | Z | x | y | z |
|---|---|---|---|---|---|
| L Superior Frontal Gyrus | 8588 | 14.3308 | -4 | 46 | 36 |
| L Frontal Orbital Cortex | 8588 | 10.6975 | -36 | 30 | -8 |
| Middle Frontal Gyrus | 8588 | 10.0424 | -36 | 18 | 44 |
| L Angular Gyrus | 2125 | 11.8496 | -40 | -64 | 36 |
| Cingulate Gyrus posterior division | 3278 | 10.8963 | 2 | -34 | 32 |
| Precuneous Cortex | 3278 | 10.7219 | 0 | -60 | 30 |
| Precuneous Cortex | 3278 | 10.393 | 2 | -50 | 12 |
| R Cerebellum IX | 299 | 9.5653 | 10 | -50 | -44 |
| R Cerebellum Crus I | 1974 | 9.2105 | 24 | -72 | -30 |
| R Cerebellum Crus I | 1974 | 5.5495 | 44 | -78 | -28 |
| L Temporal Fusiform Cortex posterior division | 635 | 9.0898 | -40 | -42 | -14 |
| R Angular Gyrus | 305 | 8.1312 | 50 | -60 | 32 |
| L Caudate | 171 | 7.1944 | -10 | 16 | 8 |
| R Frontal Orbital Cortex | 128 | 6.4057 | 38 | 36 | -10 |
| L Cerebellum Crus II | 122 | 6.227 | -42 | -64 | -40 |
| L Temporal Pole | 123 | 6.1451 | -48 | 4 | -32 |
| R Caudate | 58 | 6.0112 | 16 | 16 | 8 |
| L Cerebellum Crus I | 17 | 5.7531 | -8 | -80 | -40 |
| R Frontal Pole | 33 | 5.7378 | 32 | 60 | 4 |
| L Thalamus | 24 | 5.6025 | -2 | -2 | 6 |
| R Middle Frontal Gyrus | 34 | 5.5417 | 40 | 24 | 46 |
| L Frontal Orbital Cortex | 17 | 5.4899 | -26 | 16 | -24 |

# B

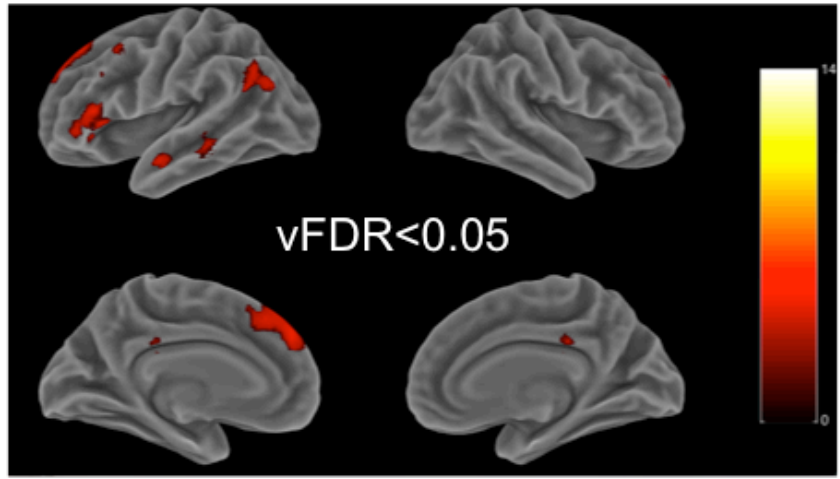

| Region | Cluster size | Peak | x | y |
|---|---|---|---|---|
| R Cerebellum | 469 | 5.1981 | 10 | -76 |
| L Superior Frontal Gyrus | 1033 | 4.9037 | -2 | 40 |
| L Frontal Pole | 1033 | 3.7747 | -4 | 56 |
| L Superior Frontal Gyrus | 1033 | 3.5821 | -6 | 20 |
| L Inferior Frontal Gyrus pars triangularis | 205 | 4.7172 | -48 | 28 |
| R Cerebellum IX | 228 | 4.4733 | 10 | -52 |
| L Middle Temporal Gyrus anterior division | 51 | 4.409 | -52 | -8 |
| L Angular Gyrus | 235 | 4.4082 | -44 | -64 |
| R Frontal Pole | 30 | 4.0744 | 16 | 54 |
| R Angular Gyrus | 11 | 4.0334 | 54 | -68 |
| Cingulate Gyrus posterior division | 29 | 3.9907 | 0 | -32 |
| R Frontal Pole | 14 | 3.9672 | 20 | 42 |
| L Middle Temporal Gyrus posterior division | 61 | 3.9528 | -60 | -32 |
| R Cerebellum Crus II | 53 | 3.8702 | 42 | -66 |
| R Temporal Fusiform Cortex posterior division | 5 | 3.7833 | 40 | -24 |
| L Middle Frontal Gyrus | 11 | 3.7053 | -36 | 14 |
| L Middle Frontal Gyrus | 6 | 3.6262 | -50 | 22 |

**Supplementary Figure 4.** Voxel-wise false discovery rate (vFDR) thresholded semantic decision activation maps for controls (A, vFDR<0.0005) and patients (B, vFDR<0.05) are shown along with tables detailing cluster and peak statistics. These data are intended to (1) demonstrate that the overall patterns of activation are robust even at stringent voxel-wise thresholds, and (2) provide more precise information about the peak locations for each group than the more liberally thresholded maps utilized in the primary analyses.

**Supplementary Analysis 5. Assessment of damage to "core conceptual network" that showed positive correlations between activity and all three language measures**.

To assess whether or not differences in activation of the core conceptual network that showed positive relationships to all three language measures might be driven by frequent damage to the network, we assessed lesion frequencies for each voxel included in the network. In addition, we assessed the relationship between mean activation within the network and total damage to the network. The results of these additional analyses are shown in Supplementary Figure 5. Only the left angular gyrus was lesioned in greater than 4 (~10% of) patients, and the voxel with the highest lesion frequency was damaged in 25/43 patients. Activation of the network was weakly negatively (not significantly) related to the amount of damage sustained by the network.

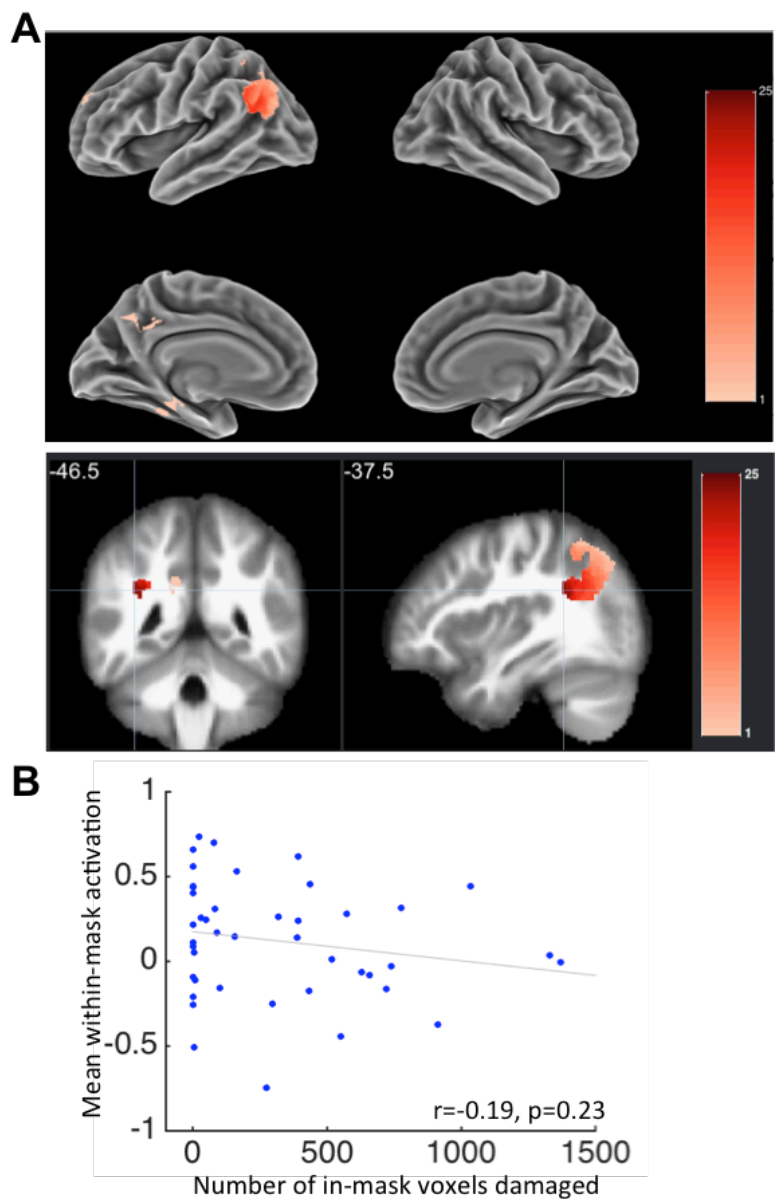

**Supplementary Figure 5. A.** Lesion frequencies for regions where activation showed significant positive relationships to all language measures (i.e. network shown in Figure 3B of the main text) are shown on a partially inflated brain (top), and on slices of a template brain with cross-hairs centered on the voxel with maximum in-mask lesion frequency. Colorbar values indicate the number of patients lesioned at each voxel. Only the left angular gyrus was lesioned in at least 10% (4/43) of patients, with slightly greater than half of patients (25/43) having damage to the left cerebral white matter underlying this region. **B.** Mean activation within the network showed a negative but non-significant relationship to the extent of damage to the network. This suggests that low levels of activity within this network are not likely to reflect widespread damage to these regions, but rather may reflect the effects of focal damage to regions that were more frequently damaged (such as the left AG) or to inter-regional connections.

**References for results from literature search shown in Supplementary Table 1.**

Abel S, Weiller C, Huber W, Willmes K (2014): Neural underpinnings for model-oriented therapy of aphasic word production. Neuropsychologia 57:154–165.

Abel S, Weiller C, Huber W, Willmes K, Specht K (2015): Therapy-induced brain reorganization patterns in aphasia. Brain 138:1097–112.

Abutalebi J, Rosa PA Della, Tettamanti M, Green DW, Cappa SF (2009): Bilingual aphasia and language control: A follow-up fMRI and intrinsic connectivity study. Brain Lang 109:141–156.

Allendorfer J, Kissela B, Holland S, Szaflarski J (2012): Different patterns of language activation in post-stroke aphasia are detected by overt and covert versions of the verb generation fMRI task. Med Sci Monit Monit 18.

Belin P, Van Eeckhout P, Zilbovicius M, Remy P, François C, Guillaume S, Chain F, Rancurel G, Samson Y (1996): Recovery from nonfluent aphasia after melodic intonation therapy: a PET study. Neurology 47:1504–1511.

Blank SC, Bird H, Turkheimer F, Wise RJS (2003): Speech production after stroke: The role of the right pars opercularis. Ann Neurol 54:310–320.

Blasi V, Young AC, Tansy AP, Petersen SE, Snyder AZ, Corbetta M (2002): Word retrieval learning modulates right frontal cortex in patients with left frontal damage. Neuron 36:159–170.

De Boissezon X, Démonet JF, Puel M, Marie N, Raboyeau G, Albucher JF, Chollet F, Cardebat D (2005): Subcortical aphasia: A longitudinal PET study. Stroke 36:1467–1473.

Bonakdarpour B, Beeson PM, DeMarco AT, Rapcsak SZ (2015): Variability in blood oxygen level dependent (BOLD) signal in patients with stroke-induced and primary progressive aphasia. NeuroImage Clin 8:87–94.

Bonakdarpour B, Parrish TB, Thompson CK (2007): Hemodynamic response function in patients with stroke-induced aphasia: implications for fMRI data analysis. Neuroimage 36:322–31.

Breier JI, Juranek J, Maher LM, Schmadeke S, Men D, Papanicolaou AC (2009): Behavioral and Neurophysiologic Response to Therapy for Chronic Aphasia. Arch Phys Med Rehabil 90:2026–2033.

Brownsett SLE, Warren JE, Geranmayeh F, Woodhead Z, Leech R, Wise RJS (2014):

Cognitive control and its impact on recovery from aphasic stroke. Brain 137:242–54.

Cappa SF, Perani D, Grassi F, Bressi F, Alberoni M, Franceschi M, Bettinardi M, Todde M, Fazio M (1997): A PET Follow-up Study of Recovery after Stroke in Acute Aphasics. Brain Lang 56:55–67.

Cardebat D, Demonet JF, De Boissezon X, Marie N, Mari RM, Lambert J, Baron JC, Puel M (2003): Behavioral and Neurofunctional Changes over Time in Healthy and Aphasic Subjects: A PET Language Activation Study. Stroke 34:2900–2906.

Connor LT, DeShazo Braby T, Snyder AZ, Lewis C, Blasi V, Corbetta M (2006): Cerebellar activity switches hemispheres with cerebral recovery in aphasia. Neuropsychologia 44:171–7.

Crinion JT, Warburton EA, Lambon-Ralph MA, Howard D, Wise RJS (2006): Listening to narrative speech after aphasic stroke: The role of the left anterior temporal lobe. Cereb Cortex 16:1116–1125.

Crinion J, Price CJ (2005): Right anterior superior temporal activation predicts auditory sentence comprehension following aphasic stroke. Brain 128:2858–71.

Crosson B, Moore AB, Gopinath K, White KD, Wierenga CE, Gaiefsky ME, Fabrizio KS, Peck KK, Soltysik D, Milsted C, Briggs RW, Conway TW, Gonzalez Rothi LJ (2005): Role of the right and left hemispheres in recovery of function during treatment of intention in aphasia. J Cogn Neurosci 17:392–406.

Eaton KP, Szaflarski JP, Altaye M, Ball AL, Kissela BM, Banks C, Holland SK (2008): Reliability of fMRI for studies of language in post-stroke aphasia subjects. Neuroimage 41:311–22.

Fernandez B, Cardebat D, Demonet JF, Joseph PA, Mazaux JM, Barat M, Allard M (2004): Functional MRI follow-up study of language processes in healthy subjects and during recovery in a case of aphasia. Stroke 35:2171–2176.

Fridriksson J (2010): Preservation and modulation of specific left hemisphere regions is vital for treated recovery from anomia in stroke. J Neurosci 30:11558–11564.

Fridriksson J, Baker JM, Moser D (2009): Cortical mapping of naming errors in aphasia. Hum Brain Mapp 30:2487–2498.

Fridriksson J, Bonilha L, Baker JM, Moser D, Rorden C (2010): Activity in preserved left hemisphere regions predicts anomia severity in aphasia. Cereb Cortex 20:1013–9.

Fridriksson J, Morrow-Odom L, Moser D, Fridriksson A, Baylis G (2006): Neural recruitment associated with anomia treatment in aphasia. Neuroimage 32:1403–1412.

Fridriksson J, Moser D, Bonilha L, Morrow-Odom KL, Shaw H, Fridriksson A, Baylis GC, Rorden C (2007): Neural correlates of phonological and semantic-based anomia treatment in aphasia. Neuropsychologia 45:1812–22.

Fridriksson J, Richardson JD, Fillmore P, Cai B (2012): Left hemisphere plasticity and aphasia recovery. Neuroimage 60:854–63.

Geranmayeh F, Leech R, Wise RJS (2016): Network dysfunction predicts speech production after left hemisphere stroke. Neurology 86:1296–1305.

Griffis JC, Nenert R, Allendorfer JB, Szaflarski JP (2016): Interhemispheric Plasticity following Intermittent Theta Burst Stimulation in Chronic Poststroke Aphasia. Neural Plast 2016:20–23.

van Hees S, McMahon K, Angwin A, de Zubicaray G, Read S, Copland DA (2014): A functional MRI study of the relationship between naming treatment outcomes and resting state functional connectivity in post-stroke aphasia. Hum Brain Mapp 35:3919–31.

Heiss WD, Kessler J, Thiel A, Ghaemi M, Karbe H (1999): Differential capacity of left and right hemispheric areas for compensation of poststroke aphasia. Ann Neurol 45:430–438.

Heiss WD, Karbe H, Weber-Luxenburger G, Herholz K, Kessler J, Pietrzyk U, Pawlik G (1997): Speech-induced cerebral metabolic activation reflects recovery from aphasia. J Neurol Sci 145:213–217.

Jarso S, Li M, Faria A, Davis C, Leigh R, Sebastian R, Tsapkini K, Mori S, Hillis AE (2014): Distinct mechanisms and timing of language recovery after stroke. Cogn Neuropsychol 0:1–22.

Karbe H, Thiel A, Weber-luxenburger G, Kessler J, Heiss W (1998): Brain Plasticity in Poststroke Aphasia : What Is the Contribution of the Right Hemisphere? Brain Lang 230:215–230.

Kiran S, Meier EL, Kapse KJ, Glynn PA. (2015): Changes in task-based effective connectivity in language networks following rehabilitation in post-stroke patients with aphasia. Front Hum Neurosci 9:1–20.

Kurland J, Naeser MA, Baker EH, Doron K, Martin PI, Seekins HE, Bogdan A, Renshaw P, Yurgelun-Todd D (2004): Test-retest reliability of fMRI during nonverbal semantic decisions in moderate-severe nonfluent aphasia patients. Behav Neurol 15:87–97.

Léger A, Démonet JF, Ruff S, Aithamon B, Touyeras B, Puel M, Boulanouar K, Cardebat D (2002): Neural substrates of spoken language rehabilitation in an aphasic patient: an fMRI study. Neuroimage 17:174–183.

Martin PI, Naeser MA, Doron KW, Bogdan A, Baker EH, Kurland J, Renshaw P, Yurgelun-Todd D (2005): Overt naming in aphasia studied with a functional MRI hemodynamic delay design. Neuroimage 28:194–204.

Martin PI, Naeser MA, Ho M, Doron KW, Kurland J, Kaplan J, Wang Y, Nicholas M, Baker EH, Fregni F, Pascual-Leone A (2009): Overt naming fMRI pre- and post-TMS: Two nonfluent aphasia patients, with and without improved naming post-TMS. Brain Lang 111:20–35.

Mattioli F, Ambrosi C, Mascaro L, Scarpazza C, Pasquali P, Frugoni M, Magoni M, Biagi L, Gasparotti R (2014): Early aphasia rehabilitation is associated with functional reactivation of the left inferior frontal gyrus: a pilot study. Stroke 45:545–52.

Meinzer M, Obleser J, Flaisch T, Eulitz C, Rockstroh B (2007): Recovery from aphasia as a function of language therapy in an early bilingual patient demonstrated by fMRI. Neuropsychologia 45:1247–1256.

Meinzer M, Flaisch T, Breitenstein C, Wienbruch C, Elbert T, Rockstroh B (2008): Functional re-recruitment of dysfunctional brain areas predicts language recovery in chronic aphasia. Neuroimage 39:2038–46.

Meinzer M, Flaisch T, Obleser J, Assadollahi R, Djundja D, Barthel G, Rockstroh B (2006): Brain regions essential for improved lexical access in an aged aphasic patient: a case report. BMC Neurol 6:28.

Meltzer JA, Wagage S, Ryder J, Solomon B, Braun AR (2013): Adaptive significance of right hemisphere activation in aphasic language comprehension. Neuropsychologia 51:1248–1259.
Miura K, Nakamura Y, Miura F, Yamada I, Takahashi M, Yoshikawa A, Mizobata T (1999): Functional magnetic resonance imaging to word generation task in a patient with Broca's aphasia. J Neurol 246:939–942.
Mohr B, Difrancesco S, Harrington K, Evans S, Pulvermuller F (2014): Changes of right-hemispheric activation after constraint-induced, intensive language action therapy in chronic aphasia: fMRI evidence from auditory semantic processing1. Front Hum Neurosci 8:1–15.
Musso M, Weiller C, Kiebel S, Müller SP, Bülau P, Rijntjes M (1999): Training-induced brain plasticity in aphasia. Brain 122 ( Pt 9:1781–90.
Naeser MA, Martin PI, Baker EH, Hodge SM, Sczerzenie SE, Nicholas M, Palumbo CL, Goodglass H, Wingfield A, Samaraweera R, Harris G, Baird A, Renshaw P, Yurgelun-Todd D (2004): Overt propositional speech in chronic nonfluent aphasia studied with the dynamic susceptibility contrast fMRI method. Neuroimage 22:29–41.
van Oers CA, Vink M, van Zandvoort MJE, van der Worp HB, de Haan EHF, Kappelle LJ, Ramsey NF, Dijkhuizen RM (2010): Contribution of the left and right inferior frontal gyrus in recovery from aphasia. A functional MRI study in stroke patients with preserved hemodynamic responsiveness. Neuroimage 49:885–93.
Papoutsi M, Stamatakis EA, Griffiths J, Marslen-Wilson WD, Tyler LK (2011): Is left fronto-temporal connectivity essential for syntax? Effective connectivity, tractography and performance in left-hemisphere damaged patients. Neuroimage 58:656–64.
Peck KK, Moore AB, Crosson BA, Gaiefsky M, Gopinath KS, White K, Briggs RW (2004): Functional Magnetic Resonance Imaging before and after Aphasia Therapy: Shifts in Hemodynamic Time to Peak during an Overt Language Task. Stroke 35:554–559.
Perani D, Cappa SF, Tettamanti M, Rosa M, Scifo P, Miozzo A, Basso A, Fazio F (2003): A fMRI study of word retrieval in aphasia. Brain Lang 85:357–368.
Postman-Caucheteux WA, Birn RM, Pursley RH, Butman JA, Solomon JM, Picchioni D, McArdle J, Braun AR (2010): Single-trial fMRI shows contralesional activity linked to overt naming errors in chronic aphasic patients. J Cogn Neurosci 22:1299–1318.
Pulvermüller F, Hauk O, Zohsel K, Neininger B, Mohr B (2005): Therapy-related reorganization of language in both hemispheres of patients with chronic aphasia. Neuroimage 28:481–9.
Raboyeau G, De Boissezon X, Marie N, Balduyck S, Puel M, Bézy C, Démonet JF, Cardebat D (2008): Right hemisphere activation in recovery from aphasia: Lesion effect or function recruitment? Neurology 70:290–298.
Richter M, Miltner WHR, Straube T (2008): Association between therapy outcome and right-hemispheric activation in chronic aphasia. Brain 131:1391–1401.
Robson H, Zahn R, Keidel JL, Binney RJ, Sage K, Lambon Ralph MA (2014): The anterior temporal lobes support residual comprehension in Wernicke's aphasia. Brain 137:931–43.
Rochon E, Leonard C, Burianova H, Laird L, Soros P, Graham S, Grady C (2010): Neural changes after phonological treatment for anomia: An fMRI study. Brain

Lang 114:164–179.

Rosen HJ, Petersen SE, Linenweber MR, Snyder AZ, White DA, Chapman L, Dromerick a W, Fiez JA, Corbetta MD (2000): Neural correlates of recovery from aphasia after damage to left inferior frontal cortex. Neurology 55:1883–1894.

Sakatani K, Xie Y, Wemara L, Li S, Zuo H (1998): Language-Activated Cerebral Blood Oxygenation and Hemodynamic Changes of the Left Prefrontal Cortex in Poststroke Aphasic Patients. Stroke 29:12–14.

Saur D, Lange R, Baumgaertner A, Schraknepper V, Willmes K, Rijntjes M, Weiller C (2006): Dynamics of language reorganization after stroke. Brain 129:1371–84.

Saur D, Ronneberger O, Kümmerer D, Mader I, Weiller C, Klöppel S (2010): Early functional magnetic resonance imaging activations predict language outcome after stroke. Brain 133:1252–1264.

Seghier ML, Bagdasaryan J, Jung DE, Price CJ (2014): The Importance of Premotor Cortex for Supporting Speech Production after Left Capsular-Putaminal Damage. J Neurosci 34:14338–14348.

Sharp DJ, Turkheimer FE, Bose SK, Scott SK, Wise RJS (2010): Increased frontoparietal integration after stroke and cognitive recovery. Ann Neurol 68:753–6.

Sims JA, Kapse K, Glynn P, Sandberg C, Tripodis Y, Kiran S (2016): The relationships between the amount of spared tissue, percent signal change, and accuracy in semantic processing in aphasia. Neuropsychologia 84:113–126.

Specht K, Zahn R, Willmes K, Weis S, Holtel C, Krause BJ, Herzog H, Huber W (2009): Joint independent component analysis of structural and functional images reveals complex patterns of functional reorganisation in stroke aphasia. Neuroimage 47:2057–2063.

Spironelli C, Angrilli A (2015): Brain plasticity in aphasic patients: intra- and inter-hemispheric reorganisation of the whole linguistic network probed by N150 and N350 components. Sci Rep 5:12541.

Szaflarski JP, Eaton K, Ball AL, Banks C, Vannest J, Allendorfer JB, Page S, Holland SK (2011a): Poststroke aphasia recovery assessed with functional magnetic resonance imaging and a picture identification task. J Stroke Cerebrovasc Dis 20:336–345.

Szaflarski JP, Allendorfer JB, Banks C, Vannest J, Holland SK (2013): Recovered vs. not-recovered from post-stroke aphasia: the contributions from the dominant and non-dominant hemispheres. Restor Neurol Neurosci 31:347–60.

Szaflarski JP, Vannest J, Wu SW, DiFrancesco MW, Banks C, Gilbert DL (2011b): Excitatory repetitive transcranial magnetic stimulation induces improvements in chronic post-stroke aphasia. Med Sci Monit 17:CR132-9.

Thiel A, Hartmann A, Rubi-Fessen I, Anglade C, Kracht L, Weiduschat N, Kessler J, Rommel T, Heiss W-D (2013): Effects of noninvasive brain stimulation on language networks and recovery in early poststroke aphasia. Stroke 44:2240–6.

Thomas C, Altenmuller E, Marckmann G, Kahrs J, Dichgans J (1997): Language processing in aphasia: Changes in lateralization patterns during recovery reflect cerebral plasticity in adults. Electroencephalogr Clin Neurophysiol 102:86–97.

Thompson CK, Riley EA, den Ouden DB, Meltzer-Asscher A, Lukic S (2013): Training verb argument structure production in agrammatic aphasia: Behavioral and neural recovery patterns. Cortex 49:2358–2376.

Thompson CK, Bonakdarpour B, Fix SF (2010): Neural mechanisms of verb argument structure processing in agrammatic aphasic and healthy age-matched listeners. J Cogn Neurosci 22:1993–2011.

Thompson CK, den Ouden DB, Bonakdarpour B, Garibaldi K, Parrish TB (2010): Neural plasticity and treatment-induced recovery of sentence processing in agrammatism. Neuropsychologia 48:3211–3227.

Tyler LK, Marslen-Wilson WD, Randall B, Wright P, Devereux BJ, Zhuang J, Papoutsi M, Stamatakis EA. (2011): Left inferior frontal cortex and syntax: Function, structure and behaviour in patients with left hemisphere damage. Brain 134:415–431.

Vitali P, Abutalebi J, Tettamanti M, Danna M, Ansaldo A-I, Perani D, Joanette Y, Cappa SF (2007): Training-induced brain remapping in chronic aphasia: a pilot study. Neurorehabil Neural Repair 21:152–60.

Warburton E, Price CJ, Swinburn K, Wise RJ (1999): Mechanisms of recovery from aphasia: evidence from positron emission tomography studies. J Neurol Neurosurg Psychiatry 66:155–61.

Warren JE, Crinion JT, Lambon Ralph MA, Wise RJS (2009): Anterior temporal lobe connectivity correlates with functional outcome after aphasic stroke. Brain 132:3428–42.

Weiduschat N, Thiel A, Rubi-Fessen I, Hartmann A, Kessler J, Merl P, Kracht L, Rommel T, Heiss WD (2011): Effects of repetitive transcranial magnetic stimulation in aphasic stroke: a randomized controlled pilot study. Stroke 42:409–15.

Weiller C, Isensee C, Rijntjes M, Huber W, Müller S, Bier D, Dutschka K, Woods RP, Noth J, Diener HC (1995): Recovery from Wernicke's aphasia: A positron emission tomographic study. Ann Neurol 37:723–732.

Winhuisen L, Thiel A, Schumacher B, Kessler J, Rudolf J, Haupt WF, Heiss WD (2005): Role of the contralateral inferior frontal gyrus in recovery of language function in poststroke aphasia: a combined repetitive transcranial magnetic stimulation and positron emission tomography study. Stroke 36:1759–63.

Winhuisen L, Thiel A, Schumacher B, Kessler J, Rudolf J, Haupt WF, Heiss WD (2007): The right inferior frontal gyrus and poststroke aphasia: a follow-up investigation. Stroke 38:1286–92.

Zahn R, Drews E, Specht K, Kemeny S, Reith W, Willmes K, Schwarz M, Huber W (2004): Recovery of semantic word processing in global aphasia: A functional MRI study. Cogn Brain Res 18:322–336.

Zhu D, Chang J, Freeman S, Tan Z, Xiao J, Gao Y, Kong J (2014): Changes of functional connectivity in the left frontoparietal network following aphasic stroke. Front Behav Neurosci 8:167.